\documentclass[conference,compsoc]{IEEEtran}

\usepackage{cite}
\usepackage{amsmath,amssymb,amsfonts}
\usepackage{algorithm}
\usepackage{algorithmic}
\usepackage{graphicx,subfig}
\usepackage{booktabs}
\usepackage{multirow}
\usepackage{textcomp}
\usepackage{xcolor}
\usepackage[hyphens]{url}
\usepackage{flushend}
\usepackage{color}
\usepackage{colortbl}
\usepackage{float}  
\usepackage[most]{tcolorbox}
\usepackage{pifont,wasysym}
\usepackage{threeparttable}
\usepackage{titlesec}
\usepackage{soul}

\usepackage[labelfont=bf]{caption}

\newlength{\tempheight}
\newlength{\tempwidth}
    
\newcommand{\rowname}[1]%
{\rotatebox{90}{\makebox[\tempheight][c]{\textbf{#1}}}}

\newcommand{\columnname}[1]%
{\makebox[\tempwidth][c]{\textbf{#1}}}

\newif\ifcomment
\commenttrue
\ifcomment
	\newcommand\edc[1]{\textbf{\textcolor{blue}{EDC: #1}}	}
	\newcommand\mn[1]{\textbf{\textcolor{orange}{Mohammad: #1}}	}
	\newcommand\yf[1]{\textbf{\textcolor{cyan}{Yufei: #1}}	}
	
\else
	\newcommand\edc[1]{}
	\newcommand\mn[1]{}
	\newcommand\yf[1]{}
\fi

\usepackage[square,sort,comma,numbers]{natbib}

\usepackage[bookmarks=true, bookmarksnumbered=true, colorlinks=true, linkcolor=linkcol, citecolor=citecol, urlcolor=urlcol, hypertexnames=true]{hyperref}

\definecolor{darkgreen}{RGB}{0, 100, 0}
\definecolor{linkcol}{rgb}{0.3,0,0}
\definecolor{citecol}{rgb}{0.3,0,0}
\definecolor{urlcol}{rgb}{0.3,0,0}
\definecolor{vlightgray}{gray}{0.925}

\IEEEoverridecommandlockouts
\usepackage{marvosym}
\usepackage[marginal,hang,stable]{footmisc}
\renewcommand{\footnoterule}{
	\kern -3pt
	\hrule width 1in
	\kern 2pt
}

\def\BibTeX{{\rm B\kern-.05em{\sc i\kern-.025em b}\kern-.08em
    T\kern-.1667em\lower.7ex\hbox{E}\kern-.125emX}}

\begin{document}
%
\title{Rethinking the Vulnerabilities of Face Recognition Systems: \\
From a Practical Perspective
}




\author{\IEEEauthorblockN{Jiahao Chen\IEEEauthorrefmark{2},
Zhiqiang Shen\thanks{* Jiahao Chen and Zhiqiang Shen contribute equally to this work. Shouling Ji is the corresponding author.}\IEEEauthorrefmark{2},
Yuwen Pu\IEEEauthorrefmark{2}, 
Chunyi Zhou\IEEEauthorrefmark{2},
Changjiang Li\IEEEauthorrefmark{3},\\
Jiliang Li\IEEEauthorrefmark{4},
Ting Wang\IEEEauthorrefmark{3}
and Shouling Ji\IEEEauthorrefmark{2}\textsuperscript{\Letter}
}
\IEEEauthorblockA{\IEEEauthorrefmark{2}Zhejiang University,\IEEEauthorrefmark{3}Stony Brook University,\IEEEauthorrefmark{4}Xi'an Jiaotong University \\
Emails: \{xaddwell, zhiqiang.shen, yw.pu\}@zju.edu.cn, zhouchunyi@njust.edu.cn, \\ meet.cjli@gmail.com, jiliang.li@xjtu.edu.cn,inbox.ting@gmail.com, sji@zju.edu.cn
}
}

\maketitle


\begin{abstract}
Face Recognition Systems (FRS) have increasingly integrated into critical applications, including surveillance and user authentication, highlighting their pivotal role in modern security systems. Recent studies have revealed vulnerabilities in FRS to adversarial (e.g., adversarial patch attacks) and backdoor attacks (e.g., training data poisoning), raising significant concerns about their reliability and trustworthiness. However, previous studies have been based on the overclaimed adversary's capabilities, with limited feasibility in practice. 

Correspondingly, in this paper, we delve into the inherent vulnerabilities in FRS through user studies and preliminary explorations. By exploiting these vulnerabilities, we identify a novel attack, \underline{F}acial \underline{I}dentity \underline{B}ackdoor \underline{A}ttack dubbed FIBA, which unveils a potentially more devastating threat against FRS: an enrollment-stage backdoor attack. FIBA enable broad-scale disruption by allowing any attacker donning a specific trigger to bypass these systems. 
This implies that after a single, poisoned example is enrolled into the database, the corresponding trigger becomes a universal key for any attackers to spoof the FRS. This strategy essentially challenges the conventional attacks by initiating at the enrollment stage, dramatically transforming the threat landscape by poisoning the feature database rather than the training data. We conduct extensive evaluations encompassing digital and physical experiments across 6 face recognition models, 5 commercial face authentication APIs, and 3 IoT devices to demonstrate FIBA's priorities over traditional attacks. Across these scenarios, FIBA achieves an attack success rate peaking at 100\%, underscoring its potential for widespread exploitation. Additionally, the impacts of various physical factors are also discussed and evaluated. Experimental findings together with our suggestions are reported to affected vendors. We conclude by proposing initial mitigation strategies for service vendors and offer a framework for preliminary defense against FIBA with evaluation results.



\end{abstract}

%
\IEEEpeerreviewmaketitle

\section{Introduction}
\label{sec:introduction}
Face recognition systems (FRS) are computer applications that identify or verify a person's identity from facial images or video frames by analyzing the facial feature patterns \cite{zhao2003face, kamgar2011toward}. These systems have become increasingly important across multiple domains owing to their efficient and non-invasive identification and authentication capabilities. For instance, between 2017 and 2019, 64 countries adopted artificial intelligence surveillance through FRS~\cite{32FRStatistics} and 7 in 10 governments use FRS extensively, among which 20\% use it on some buses, 30\% use it on trains or subways, and about 60\% use it in some airports~\cite{32FRStatistics}.
Recent researches~\cite{wuuniid, zhong2020towards, komkov2021advhat} have demonstrated that these security-sensitive systems are vulnerable to adversarial patch attacks achieved through elaborate patches or other disguises during the authentication. Traditional backdoor attacks~\cite{xue2021backdoors, wenger2021backdoor} can backdoor FRS by poisoning the training data of the face recognition models in advance. However, these attacks are founded on assumptions that attackers are capable of generating different patches with intensive computation resources for each source-target pair or manipulating the training data. These assumptions result in large gaps of adversary's capabilities between research and practice~\cite{damer2018morgan,scherhag2017vulnerability,taskiran2020face}, substantially increasing the cost for adversaries in real-world scenarios. 


\begin{figure}
    \centering
    \includegraphics[width=0.7\linewidth]{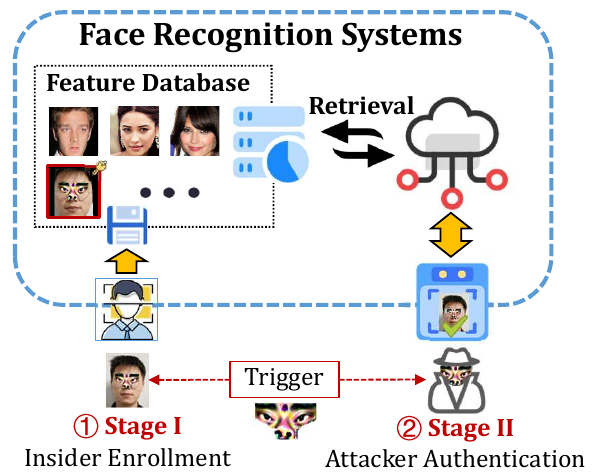}
    \caption{Illustration of FIBA. An insider wearing a trigger registers his/her face into the database. During authentication, attackers with the same trigger can bypass FRS.
    }
    \label{fig:illustration}
\end{figure}

From the perspective of attackers, we wonder that ``\ul{\textit{If it is possible to spoof a targeted FRS with practical adversary's capabilities}}". Under this practical scenario, details of the FRS (e.g., training data of the feature extractor) are out of reach, launching resource-intensive attacks is impractical and so on. Besides, considering the adversary's goal, attack performance (e.g., applicable to any attackers, stealthy to evade detection, generalizable to different FRS and robust in physical world) under complex scenarios is also important. As such, executing such an attack is far from trivial. To achieve the mentioned attack goals and further gain more insights into the vulnerabilities inherent in FRS, we first conducted a user study, with 80 volunteers involved, focused on naturally occurring impersonation and evasion attacks in real-world scenarios. The results indicate that, despite claims of accuracy exceeding 99\% by numerous facial recognition service vendors, 47.95\% and 80.82\% of FRS users have experienced misidentification or outright rejection by these services, respectively. After that, to figure it out, we conduct preliminary experiments and reveal the inherent flaws of FRS, that face recognition models rely on a small subset of features to distinguish two facial images of different identities. By exploiting these flaws, we then introduce an enrollment-stage \textbf{F}acial \textbf{I}dentity \textbf{B}ackdoor \textbf{A}ttack (FIBA)\---a novel attack vector that allows attackers to bypass FRS by enrolling an insider's face disguised with a trigger into the system, without the requirement for training data accessibility. Attackers possessing the trigger can subsequently fool the FRS with ease. \ul{\textit{Note that an insider denotes the one who has the ability to register his/her face into the database, a common privilege in FRS}}~\cite{Invixium,HikVision} (see in Section~\ref{sec:threat}). The overall attack process is shown in Figure \ref{fig:illustration}. The underlying mechanism is that distinguished facial features can be substituted with our meticulously crafted features to serve as a backdoor. Faces disguised with the same trigger will represent similar facial features. 

Even with the help of a compromised insider to register a facial image into the database of FRS, launching FIBA in real-world scenarios is still challenging since the adversaries have minor knowledge about FRS, e.g., inaccessibility of the pre-trained models of commercial APIs or IoT devices. In such a black-box setting, ensuring the generalizability of FIBA across different FRS presents significant challenges. Furthermore, bypassing FRS also entails evading facial liveness verification (FLV), a task that is likewise nontrivial. Additionally, different from previous attacks \cite{wuuniid,li2020light}, the attacker's images remain inaccessible, necessitating its universality towards various identities. The robustness of FIBA is also vital for resisting the effects of digital-to-physical distortion that may degrade the attack performance. Consequently, it is imperative to address the following challenges: 
\begin{enumerate}
    \item \textbf{Generalization:} \textit{How to eliminate the gap between black-box and white-box settings in practice?}
    \item \textbf{Stealthiness:} \textit{How to maintain the stealthiness of FIBA to evade the detection of FLV and its effectiveness simultaneously?}
    \item \textbf{Universality:} \textit{How to make it universal to any attackers at once without their facial images?}
    \item \textbf{Robustness:} \textit{How to make it robust to digital-to-physical distortion to keep attack performance?}
\end{enumerate}

To address challenge 1), a data-dependent backdoor feature optimization algorithm is introduced to eliminate the parameter disparities and other differences among the facial feature extractors. This algorithm leverages the common vulnerabilities within FRS to optimize the generation of the patch (trigger) with key facial features, effectively mitigating disparities caused by model parameters and structures. To safeguard the attack's stealthiness and effectiveness in challenge 2), exploratory experiments are conducted to pinpoint the locations of key facial features in advance, ensuring the minimization of the facial alteration area while maximizing attack performance. To fulfill the universality in challenge 3), attacker-agnostic patch optimization flow is used by applying the same patch for both attackers and insiders. Finally, to address challenge 4), external dataset augmentation and physical transformation techniques are employed to guarantee the robustness of generated triggers to digital-to-physical distortion. With the above challenges addressed, FIBA can breach practical applications easily. The following is an example attack scenario: for a university deployed with FRS, a student can enroll his/her disguised face and distribute the trigger for profit (note that enrolling or updating facial images are common in practice, e.g., HikVision \cite{HikVision} and Invixium \cite{Invixium}), allowing anyone with the same trigger to bypass the FRS. Since the trigger is applicable to anyone, this threat can be amplified rapidly, as the trigger is spread to more people. FIBA can also introduce similar threats in companies, housing estates, and other departments equipped with FRS (i.e., imagine that everyone could bypass the FRS of a security-critical department).

We validated the universality and generalization of FIBA through extensive experiments. These experiments encompassed simulations across four datasets involving six face recognition models, evaluations using five mainstream commercial APIs, and experiments utilizing three IoT edge devices. With these experimental results and analysis, our work has systematically uncovered the enrollment-stage vulnerabilities (when there are malicious attackers) of FRS for the first time. In summary, the key contributions of this paper can be summarized as follows:
 \begin{itemize}
     \item We reveal the inherent vulnerabilities within FRS through user studies and subsequently analyze the underlying causes via a preliminary experiment.
     \item From a practical perspective, we propose FIBA, an enrollment-stage backdoor attack enabling attackers to spoof FRS and evade FLV with a compromised insider's registered identity.
     \item Experiments on 6 face recognition models, 5 mainstream commercial APIs, and 3 IoT edge devices are conducted. A maximum attack success rate of 100\% is achieved under different settings in the physical domain. IoT devices with FLV can also be cheated with a maximum attack success rate of 100\%.
     \item Based on our analysis, we have reported our experimental findings to affected vendors together with our suggestions. A potential defense strategy is proposed to mitigate the threat of FIBA, and a preliminary evaluation is conducted to shed light on building more reliable FRS.
 \end{itemize}
\textbf{Ethics.} All user studies are completed by anonymous participation of volunteers to avoid sensitive data disclosure. Approval of all ethical and experimental procedures and protocols was granted by Science and Technology Ethics Committee of
XXX University.

\begin{table*}[t]
\centering
\caption{Comparison of different attacks against FRS.}
\label{tab:compare_method}
\begin{threeparttable}
\begin{tabular}{ccccccc}
\toprule
Attack Type & Poison Stage & Feasibility & Universality & Generalization & Stealthiness & Robustness \\ 
\midrule
Morphing Attack& Enrollment & \CIRCLE & \Circle & \LEFTcircle & \Circle & \Circle \\
Adversarial Attack & \-- & \LEFTcircle & \LEFTcircle & \LEFTcircle & \CIRCLE & \CIRCLE \\
Backdoor Attack & Training & \Circle & \CIRCLE & \Circle & \CIRCLE & \Circle\\
FIBA (\textbf{Ours}) & Enrollment & \CIRCLE & \CIRCLE & \CIRCLE & \CIRCLE & \CIRCLE\\ 
\bottomrule
\end{tabular}
\begin{tablenotes}
\footnotesize
\item [1] \CIRCLE, \LEFTcircle, \Circle, from high to low, represent different matching degrees; ``-" means no requirement.
\end{tablenotes}
\end{threeparttable}
\end{table*}

\section{RELATED WORK}
\label{sec:background}
This section primarily discusses various attacks targeting FRS and highlights their limitations in real-world scenarios.

\subsection{Face Morphing Attacks} 
Face morphing attack \cite{venkatesh2021face} involves creating a digitally composite or ``morphed" face image. The ``morphed" face combines the facial features of two or more individuals of different identities, allowing the attackers to bypass authentication processes after enrolling the generated facial image into the database. Morphing attacks can be conducted in either the image or feature space. The former generates a forged image by interpolating the facial images of different identities together \cite{scherhag2017vulnerability, wolberg1998image}. The latter aims to reconstruct an average face to make it close to the given identities in feature space, using a generative adversarial network \cite{damer2018morgan, scherhag2017vulnerability} for generating morphed face. However, the performance of these attacks is limited by the number of identities that need to be fused. For example, Wu et al. \cite{wuuniid} introduced UniID, a method to inject a specialized disguised face during the enrollment phase, enabling multiple attackers to bypass the FRS. Since UniID necessitates the facial images of the attackers to generate the patch to disguise the enrolled face, its performance diminishes nearly to zero as the number of attackers increases to 10. Li et al. \cite{li2020light} presented an attack that directly injected the backdoor by modulating LED in a specialized waveform during the enrollment phase, raising the assumption for adversaries' capabilities. Besides, their method requires accessibility to facial images of the attackers, which can thus be viewed as a morphing attack.

\subsection{Adversarial Attacks against FRS}
The advancements of adversarial learning have also opened the door for potential attackers to spoof by face forging or wearing carefully crafted stickers or constrained adversarial patch $p$:
\begin{equation}
    \arg\min\limits_{p} l(f_{\theta}(x_a+p),f_{\theta}(x_t))
\end{equation}
where $f_{\theta}(x_t)$ is the feature vectors of targeted identity for optimization and $l(\cdot)$ is the overall loss function to satisfy attack constraints. This goal implies that for each pair of attacker and victim $(x_a,x_t)$, the attacker must generate a distinctive patch $p$, entailing a significant attack cost \cite{komkov2021advhat, jia2022adv}. The concept of a universal adversarial attack has been proposed to enable adversarial perturbations to be universally applicable across attackers with different identities. Amada et al.  \cite{amada2021universal} utilized small, pixel-level perturbations that, when added to facial images, can deceive the FRS into recognizing a face as belonging to several distinct identities simultaneously. Yang et al. \cite{yang2020design} developed a ``universal adversarial patch" that can reliably fool face detectors into failing to detect the faces. However, these adversarial attacks are only launched in the authentication stage, and their generalization and universality are limited by the difference between the surrogate and target model.

\subsection{Backdoor Attacks against FRS}
Backdoor attacks against FRS entail manipulating the model's training data to introduce a concealed vulnerability that attackers can exploit in the authentication stage. This attack aims to compromise the system's integrity and enable unauthorized access to bypass facial recognition authentication. Guo et al. \cite{guo2021master} introduced a hidden backdoor called ``master key" into the feature extractor during its training phase and activated the backdoor using a specific trigger. These attacks \cite{guo2021master, sarkar2020facehack, xue2021backdoors} focus on compromising the FRS by poisoning the training data of the feature extractors. Nevertheless, such an endeavor is frequently unrealistic, given that access to the training data is often restricted. 

\begin{figure}[t]
    \centering
    \includegraphics[width=0.7\linewidth]{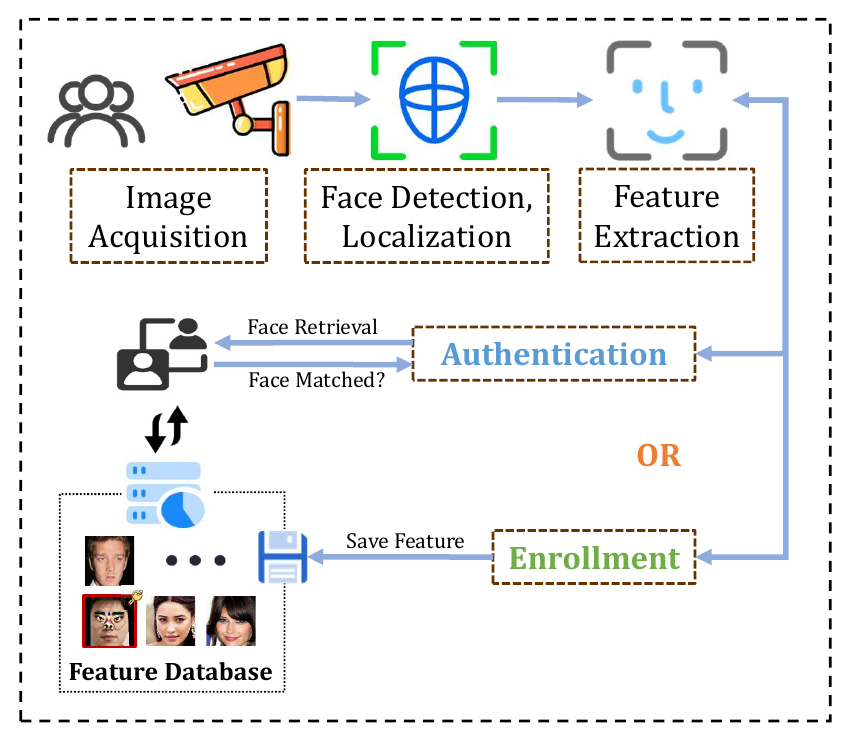}
    \caption{Workflow of the traditional FRS.}
    \label{fig:frs}
\end{figure}

\subsection{Remarks}

Systematically, we summarize our attack and list the differences between previous techniques from several dimensions that matter a lot in real-world scenarios.

\textbf{Feasibility.} Feasibility denotes whether the attacks can be launched easily in real-world scenarios. Backdoor attacks \cite{guo2021master, sarkar2020facehack, xue2021backdoors} requiring the manipulation of the training data is far from feasible since FRS are often unknown to attackers. Adversarial attacks \cite{deb2020advfaces,kurakin2018adversarial} need extra computations for each new attacker, assuming high computational resources for the adversary. As for face morphing \cite{scherhag2017vulnerability, wolberg1998image} and FIBA, the only step that needs computation is to generate the ``average or triggered face" and enroll it in the database.

\textbf{Universality.} Universality means whether the attacks are universally applicable to any attackers when attackers' facial images are out of reach. This characteristic measures the severity of the threat since high universality brings large-scale and rapid spread. Face morphing attacks \cite{scherhag2017vulnerability, wolberg1998image} that need attackers' facial images to generate morphed faces have little universality. Adversarial attacks \cite{demontis2019adversarial,huang2019enhancing,zhou2018transferable} have minor universality because of their transferability. Backdoor attacks \cite{guo2021master, sarkar2020facehack, xue2021backdoors} and FIBA do not specify attackers and are universally applicable.

\textbf{Generalization.} Generalization refers to whether the attacks are effective to different target FRS under the most practical (black-box) settings, with no accessibility to models of FRS. Since both morphing and adversarial attacks \cite{demontis2019adversarial,wolberg1998image} rely heavily on the surrogate model, their generalization is limited. For the black-box FRS, backdoor attacks \cite{guo2021master, sarkar2020facehack, xue2021backdoors} have no authority to modify the training data of the pretrained recognition models as well. Note that under the black-box setting, FIBA hardly depends on models, with high transferability which is validated in the evaluation.

\textbf{Stealthiness.} Stealthiness measures whether the attacks will be detected since advanced FRS have deployed with FLV and other anomaly detection methods. Face morphing attacks \cite{damer2018morgan,scherhag2017vulnerability} often optimize in the digital space to generate a morphed face without any stealthiness to evade FLV. However, the other three attacks \cite{zhao2019seeing,xu2020adversarial,wiyatno2019physical} can be launched with high flexibility to avoid potential detection.

\textbf{Robustness.} Robustness denotes whether the attacks will be affected by external disturbance, especially in the physical domain, where many factors may degrade the attack performance. Previous morphing \cite{damer2018morgan} and backdoor \cite{guo2021master} attacks tend to focus on performance in the digital domain while ignoring the robustness in the physical domain. Adversarial attacks \cite{zhao2019seeing,xu2020adversarial,kurakin2018adversarial} and FIBA have made many efforts to improve their robustness to launch attacks in the physical domain with high ASR.

Based on the above findings, we summarize existing attacks as follows:

\begin{tcolorbox}[title = {Remark}]
From a practical perspective, there are large gaps between these attacks and real-world scenarios since they fail to keep the balance among feasibility, universality, generalization, stealthiness and robustness. 
\end{tcolorbox}

\section{MOTIVATION}
\label{sec:motivation}
In this section, the fundamental framework and workflow of the FRS are introduced, followed by a user study on FRS conducted to obtain a comprehensive understanding of the vulnerabilities of FRS in real-world conditions.

\subsection{Face Recognition Systems}
The field of face recognition technology \cite{meng2021magface,wu2023face} has advanced rapidly, finding widespread applications across various domains, ranging from secure access control to personal device user authentication. At the core of these systems lies a multi-stage workflow that converts the input images into facial representations suitable for efficient matching and recognition. As shown in Figure \ref{fig:frs}, this workflow can be broadly divided into two key stages: face enrollment and authentication, both of which include face detection and localization, feature extraction, and face matching or saving.

The initial face detection stage aims to localize and isolate regions within the input image that potentially contain a human face and the next stage aims to localize key facial landmarks. With facial landmarks established, the system can normalize the detected face by applying geometric transformations like scaling, rotation, and alignment. Finally, having obtained these normalized face representations: (a) during the \textbf{enrollment phase}, the extracted representations will be stored in the feature database for retrieval; (b) during the \textbf{authentication phase}, the face matching stage compares the extracted representations with features in the database.

For formalization, let $f_{\theta}(x): \mathcal{X}\to\mathcal{Y}$ represents a facial feature extractor, with $x\in\mathcal{X}=[0,1]^{C\times W\times H}$ and facial features $y\in\mathcal{Y}=[0,1]^{1\times K}$. To search whether there is an identity in the database $\mathbb{D}$ means:
\begin{equation}
    A = \{\mathrm{id}|\mathrm{cos}(f_{\theta}(x),y_{\mathrm{id}})>\mathrm{\delta_{match}}, y_{id}\in\mathbb{D}\}
\end{equation}
where $x$ is a normalized face image, $\mathrm{cos}(\cdot)$ denotes cosine similarity and $\mathrm{\delta_{match}}$ is a threshold for face matching. If the result $A=\phi$, the identity does not exist in $\mathbb{D}$, and vice versa. Therefore, $f_{\theta}$ is the key component for FRS to distinguish various identities. To successfully spoof the FRS, it is crucial to deceive $f_{\theta}$ first.

\subsection{Inherent Vulnerability of FRS}
\begin{figure}[t]
    \centering
    \includegraphics[width=0.8\linewidth]{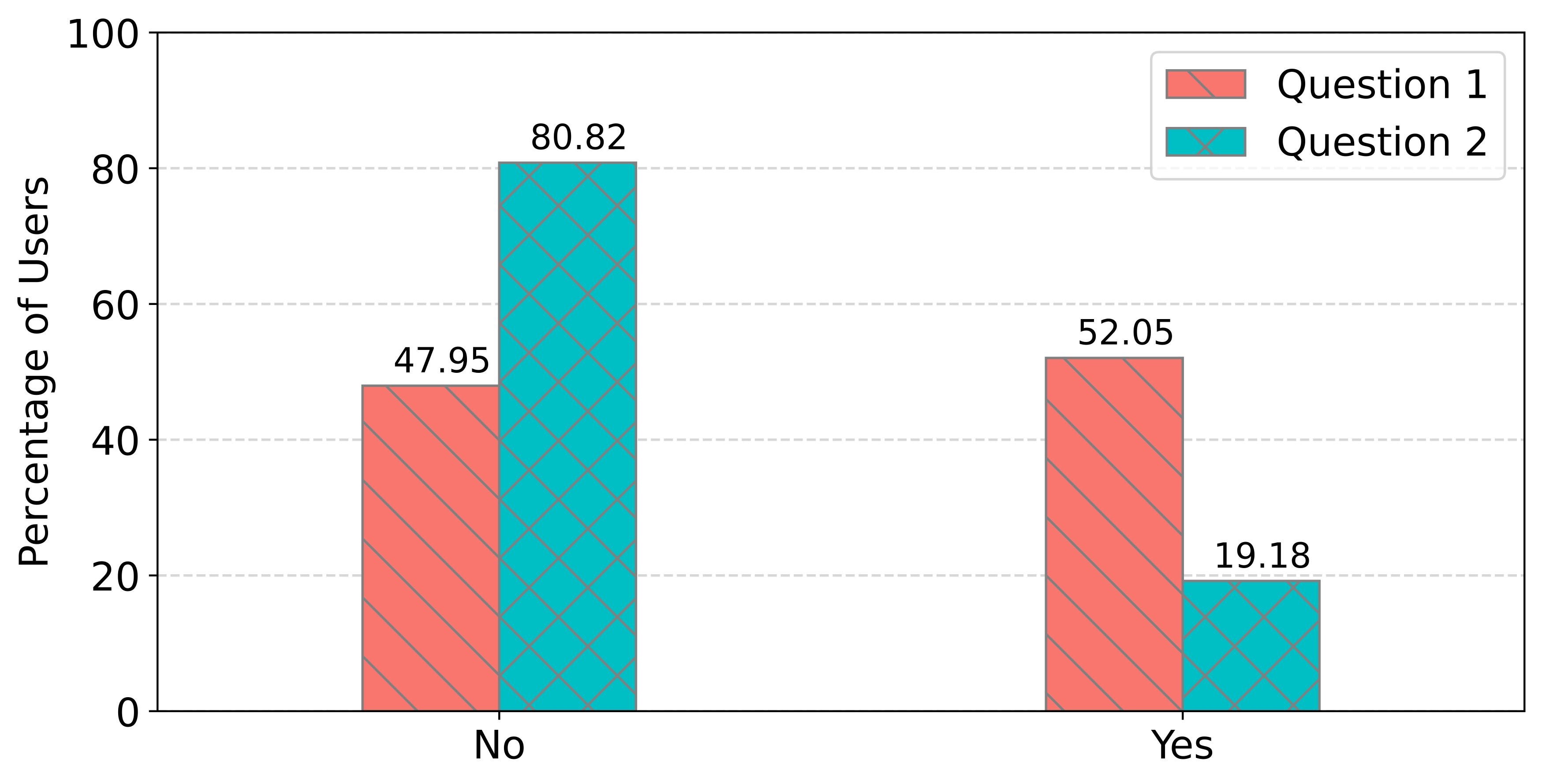}
    \caption{Results of user study. \underline{Question 1}: \textit{\textmd{Have you ever been mistaken for someone else when using FRS?}} \underline{Question 2}: \textit{\textmd{Have you ever been unrecognized when using FRS?}}}
    \label{fig:user_study}
\end{figure}

With the preliminaries of FRS, we are going to investigate the vulnerabilities in real-world settings. Starting from our own experience (often misidentified as others), we notice that FRS is susceptible not only to deliberate malicious attacks but also to various environment variables, significantly affecting user experience. To obtain a more comprehensive understanding of this phenomenon, a user survey is conducted to collect instances of mistaken identity recognition or failure to recognize users during FRS utilization. 80 volunteers (details presented in Appendix \ref{appendix:user_study}) are recruited, and each participant is administered a questionnaire. It should be noted that 73 participants have prior experience using FRS. The user survey results are depicted in Figure \ref{fig:user_study}. For those who have used FRS before, a staggering 47.95\% of participants reported instances of being misidentified as someone else, and an overwhelming 80.82\% experienced failures to be recognized by the system, highlighting that there are inherent flaws within FRS, presenting a significant obstacle to widespread adoption and seamless user experience. These naturally occurring evasion and impersonation incidents suggest that FRS is not as reliable as previously assumed, even without deliberate attacks from adversaries. These findings act as a sobering wake-up call, prompting an in-depth investigation into these issues' root causes and ramifications. A pressing question demands our attention: 

\begin{tcolorbox}[]
\textit{
What underlying factors contribute to such high rates of misidentification and recognition failures?
}
\end{tcolorbox}

\subsection{Experimental Exploration and Analysis}
To answer the above question and explore these naturally occurring attacks, we formalize the recognition process of queried user image $x_q$ as follows:
\begin{equation}
\begin{aligned}
    &\mathrm{cos}(f_{\theta}(\Phi(x_q)),e_q)\le\mathrm{\delta_{match}}\quad or\\
    &\mathrm{cos}(f_{\theta}(\Phi(x_q)),e_k)\geq\mathrm{\delta_{match}}, k\neq q
\end{aligned}
\end{equation}
where facial features $e_q,e_k\in\mathbb{D}$ and $\Phi(\cdot)$ denotes the transformation that makes $f_{\theta}(\Phi(x_q))$ different from $e_q$, leading to the misidentification. In practice, $\Phi(\cdot)$ simulates the factors introduced by the environment  (e.g., light intensity, quality of the images, etc) and human (makeup, shooting angles, expressions, decorations, etc). All of these factors work together and result in recognition errors.

\begin{figure}[t]
    \centering  
    \vspace{-0.4cm} 
    \subfloat{
        \includegraphics[width=0.26\linewidth]{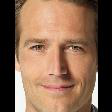}}
    \quad 
        \subfloat{
        \includegraphics[width=0.26\linewidth]{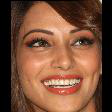}}
    \quad 
    \subfloat{
        \includegraphics[width=0.26\linewidth]{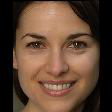}}\\
    \setcounter{subfigure}{0}
    \subfloat[similarity=29.52]{
        \includegraphics[width=0.26\linewidth]{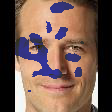}}
    \quad
        \subfloat[similarity=19.58]{
        \includegraphics[width=0.26\linewidth]{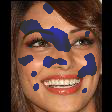}}
    \quad
    \subfloat[similarity=24.73]{
        \includegraphics[width=0.26\linewidth]{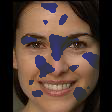}}\\
    \caption{Similarity of masked facial images. The first row of images denotes the source face of masked faces in the second row.}
    \label{fig:interpret}
\end{figure}

Nevertheless, $\Phi(\cdot)$ is not the direct cause since these factors always exist, but natural attacks seldom occur. Instead, understanding how these factors influence the feature extractor $f_{\theta}(\cdot)$ is more important. Considering that all of these changes ultimately manifest in the alteration of the feature vectors, our goal is to figure out what features matter a lot in distinguishing ourselves from others. To achieve it, we devise a feature-level algorithm to interpret the abnormal behaviors of $f_{\theta}$. Specifically, we need to find out the most important facial region (features) that determine the decisions of FRS to distinguish different faces. We define:
\begin{equation}
    x\oplus p = x\cdot(1-m)+p\cdot m
\end{equation}
where $m$ represents the mask for the given patch $p$, that is:
\begin{equation}
    m_{i,j}=\begin{cases}
    1, & \text{if } p_{i,j} \neq 0 \\
    0, & \text{otherwise}
    \end{cases}
\end{equation}
Therefore, $x\oplus p$ denotes the facial image with a patch attached. Now, we aim to generate a patch $p$ to minimize the similarity between $x$ and $x\oplus p$ given $f_{\theta}$, through which we can find the most important features that determine the decisions of FRS. Further, the facial region of the patch denotes the key features of the face. Considering that facial features are related to spatial location, we keep the patch's spatial continuity with TV loss \cite{sharif2016accessorize} and aggregation loss \cite{wei2023physically}. The primary mechanism is formalized as:
\begin{equation}
    \arg\max\limits_{p} l(x,x\oplus p,f_{\theta}), \mathrm{ s.t.} \|p\|_{n}\le \gamma
\end{equation}
where $l(\cdot)$ is the loss function for optimization. We also restrict the changing area of $p$ with $\|p\|_{n}\le \gamma$. The details of the algorithm can be found in Appendix \ref{appendix:mask}. For the sake of experimental simplicity, we only set the changing areas of $p$ to 10\% (for image space), and $p$ is randomly initialized in each iteration to avoid its influence on the optimization of $m$. The result is shown in Figure \ref{fig:interpret}, using CelebA  \cite{karras2017progressive} and MobileFace \cite{chen2018mobilefacenets} for optimization. From the similarity between $x$ and $x\oplus p$, we can indicate that some features are more essential for $f_{\theta}$ than others, and only a small part of the occlusion of these key features can blind FRS. Based on the analysis above, the unrecognition ($\mathrm{cos}(f_{\theta}(\Phi(x_q)),e_q)\le\mathrm{\delta_{match}}$) of FRS can be attributed to the influence of various factors on the key features of feature extraction. 

\begin{figure}[t]
  \centering
  \includegraphics[width=0.9\linewidth]{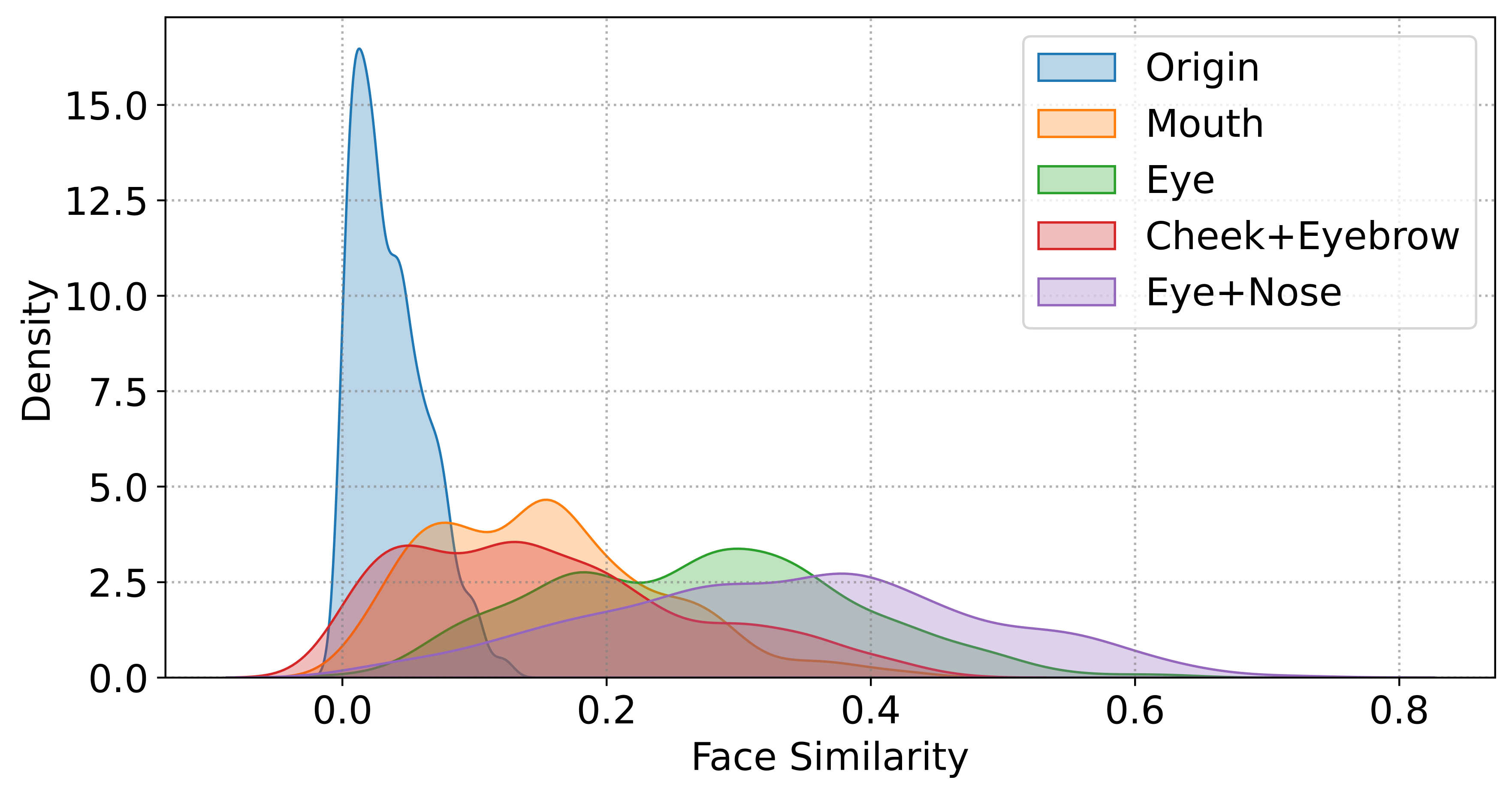}
  \caption{Similarity distribution of the different mask regions (on CelebA dataset with MobileFace). }
  \label{fig:mask_region}
\end{figure}

The above inference can also be associated with the misidentification ($\mathrm{cos}(f_{\theta}(\Phi(x_q)),e_k)\geq\mathrm{\delta_{match}}, k\neq q$) of FRS, since faces without distinguished features could easily be matched with an ``average face" which also lacks distinguished features in the database. To validate this conjecture, we randomly sample 1000 face pairs $(x_i,x_j)$ from CelebA as $\mathbb{D}_{pair}$, where the identity of $x_i$ is different from that of $x_j$. Then, we use prefabricated masks (see in Appendix \ref{appendix:mask}) $\mathbb{M}=\{m_0,m_1,...,m_i\}$ for different parts of the face. Based on the above preparations, we can draw a histogram of the similarity of $\mathrm{cos}(f_{\theta}(x_i\cdot (1-m_k)),f_{\theta}(x_j\cdot (1-m_k)))$, for $m_k\in\mathbb{M}$ and $(x_i,x_j)\in\mathbb{D}_{pair}$. Figure \ref{fig:mask_region} illustrates that the distribution of face similarity will move forward as a result of the addition of the masks, and different regions present different impacts on the similarity changes. For instance, mask regions, including the eye and nose, can largely increase the similarity of faces from different identities. However, masks of mouth or cheek have less impact on the feature extraction of FRS. With these observations, we can also arrive at a conclusion consistent with the previous experiments, which is that facial feature extractors tend to focus more on the local features of faces to distinguish between facial images from different identities. Figure \ref{fig:mask_region} also presents that when a small set of these distinctive features is removed (i.e., $x\cdot (1-m)$), the facial feature extractor struggles to discern these faces. That's to say, if a feature extractor fails to extract the distinguished features of some facial images in the enrollment and authentication stage, these images will probably have higher similarities. That's why many of us often encounter situations where we are mistakenly identified as other individuals.

\begin{figure*}[h]
\centering
    \includegraphics[width=0.6\linewidth]{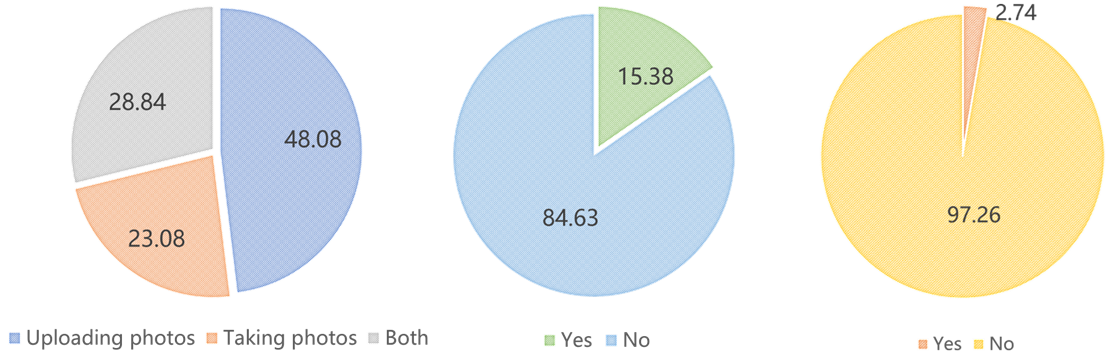}
   \caption{Results of user study. From left to right are, \underline{Question 1}:\textit{\textmd{Which type of enrollment did you use often?}} \underline{Question 2}:\textit{\textmd{Were you supervised during the enrollment?}} \underline{Question 3}:\textit{\textmd{Were you supervised during the authentication?}}}
   \label{fig:user_study2}
\end{figure*}

With the above observations and analysis, we can infer that these naturally occurring evasion and impersonation attacks occur because of the inherent vulnerabilities of FRS, and it is necessary for FRS users to avoid the occlusion of key features like eyes or nose during both the enrollment and authentication stage. Concisely, we summarize this conclusion as a takeaway:
\begin{tcolorbox}[title = {Takeaway}]
Only a small portion of the human face represents key features that distinguish us from others, without which it's easy for FRS to misidentify us.
\end{tcolorbox}

However, we also question:
\begin{enumerate}
    \item \textit{Whether it will cause serious harm to safety-critical applications if these inherent vulnerabilities are exploited by malicious adversaries?}
    \item \textit{Are the FRS in real-world scenarios robust against these deliberate attacks?}
\end{enumerate}

We address the above questions by introducing FIBA and evaluating it in the digital and physical domains. The rest of the paper proceeds as follows. The threat model of our proposed attack is presented in Section \ref{sec:threat}. We then detail the design of FIBA to compromise FRS in Section \ref{sec:method}. In Section \ref{sec:evaluation}, we assess the robustness of FRS with comprehensive experimental results encompassing both digital and physical domains, quantifying the severity of these flaws from the perspective of service vendors. Based on the evaluation analysis, we propose a potential defense method to mitigate FIBA and offer suggestions to face recognition vendors to enhance the robustness of FRS. Finally, we discuss challenges and directions for constructing reliable and trustworthy FRS.

\section{THREAT MODEL}
\label{sec:threat}
In this section, we first outline the adversary's goal related to FRS. We then explore the adversary's capabilities to launch an attack under different conditions.

\textbf{Adversary's Goal.} Note that we consider the traditional FRS architectures, which have been widely deployed in the real world. From attackers' perspective, they always seek to gain more authority to bypass safety-critical systems like FRS with less cost.

To circumvent FRS, we consider the following attack scenarios: A system insider becomes compromised and intends to allow attackers to bypass the FRS for malicious purposes with his/her own privileges (by enrolling a face into the FRS); another situation involves the insider himself desiring to use his/her privileges to make illegal profits by selling his/her authority that can bypass FRS to attackers. Both of the situations are practical~\cite{wuuniid,taskiran2020face} and would wreak havoc on FRS if they happened in the real world.

However, managing this with only an insider's endeavor is not trivial. To achieve this, the attackers can provide the insider with a specific and wearable trigger used for disguise. The insider wearing a mask can enroll his/her face into the database. After that, both the insider and attackers can bypass the FRS by wearing the same trigger without the need for extra optimization during each authentication.

\textbf{Adversary's Knowledge.} Following the settings of previous work \cite{wuuniid}, we assume that an insider may have white-box or black-box access to FRS. Thus, attackers have the same knowledge about FRS as the insider if he/she has been compromised. In the white-box settings, attackers can use a feature extractor of the target FRS for optimization. In the black-box setting, information about the FRS is out of reach; attackers can only use a surrogate model to generate a mask and launch a transfer-based attack. Additionally, we do not assume that facial images of the attackers are accessible since the attackers are unknown in the second scenario mentioned before.

\textbf{Adversary's Capabilities.} In this paper, we assume that the compromised insider has the ability to enroll an identity in FRS. What needs to be emphasized is that this privilege is common for insiders of the FRS (e.g., students can update their facial images on a specified APP easily to avoid misidentification caused by their changes in appearance \cite{HikVision,Invixium}). They can complete this step by uploading or taking photos with IoT devices, and corresponding feature vectors will be stored in the database. It should be noted that both of these methods are practical in the real world. To demonstrate this, we expanded upon the user survey results from a previous work \cite{wuuniid} and added a question (73 users have answered Question 3 with details presented in Appendix \ref{appendix:user_study}). Results shown in Figure \ref{fig:user_study2} indicate that it is common for many FRS users to enroll either digital images or photos taken by edge devices and the devices used for evaluation also validate that both approaches are used in real-world scenarios (see in \ref{appendix:devices}).
Moreover, FRS users are rarely supervised during both the enrollment and authentication phases. Thus, this paper considers two conditions: uploading facial images and taking photos. Uploading facial images directly means less digital-to-physical distortion while taking facial photos in the physical domain must consider this distortion, which will have a negative impact on FIBA.

\section{METHODOLOGY}
\label{sec:method}
As mentioned in Section \ref{sec:motivation}, feature extractor $f_{\theta}$ of FRS tends to focus on a small subset of features to distinguish two facial images of different identities. Our exploration also suggests that by masking these key features, the ability of $f_{\theta}$ is largely restricted. Inspired by these findings, we devise a novel attack to exploit this characteristic of FRS and the detailed design is presented in this section.

\subsection{Backdoor Trigger Generation}
Based on the preliminary analysis, attack objectives are proposed in the following paragraphs. Given a pretrained facial feature extractor $f_{\theta}$, an adversary intends to fool it by cooperating with a legitimate but compromised insider whose facial image $x_v$ can be enrolled in the feature database $\mathbb{D}$. Specifically, the adversary generates a wearable patch to replace the key features of the face and sends it to the insider. The insider wearing the patch then enrolls his/her face in the database. The adversary wearing the same patch can also bypass the FRS. To this end, our optimization objective can be formalized as:
\begin{equation}
    p = \arg\min\limits_{p} \mathcal{L}(p,m,x_v,f_{\theta})
\end{equation}
where $p$ and $m$ denote patch and mask, respectively. Thus, $x_v\oplus p$ refers to the enrolled face with patch $p$ (backdoor trigger). $\mathcal{L}(\cdot)$ is the overall loss to achieve adversary's goal. 

\textbf{Multi-objective Similarity.} To ensure that the facial feature vectors of an enrolled insider are as close to that of unknown triggered users as possible, we can collect some facial images from the Internet as a candidate dataset $D_{c}$ for patch optimization. Thus, we have:
\begin{equation}
\label{eq:loss_sim}
\begin{aligned}
    \mathcal{L}_{sim} = \mathbb{E}_{x_i\in D_{c}}\mathrm{cos}(\frac{f_{\theta}(x_i\oplus p)}{\|f_{\theta}(x_i\oplus p)\|_2},\frac{f_{\theta}(x_v\oplus p)}{\|f_{\theta}(x_v\oplus p)\|_2})
\end{aligned}
\end{equation}
Note that different from the previous work \cite{wuuniid}, it's easy for FIBA to maximize this term $\mathcal{L}_{sim}$ since our findings reveal that $f_{\theta}$ only focuses on key features of faces. By masking these features with $m$ and replacing them with patch $p$, this term can converge quickly.

\textbf{Color Discrepancies.} When generating a patch in the digital domain, the real-world loss should also be considered: 1) color discrepancies resulting from printing equipment; 2) inconsistency caused by environmental factors like illumination, resolutions of edge devices and image compression algorithm; 3) variations made by FRS users like gestures, makeup, expression, and so on. For color discrepancies, an existing study \cite{sharif2016accessorize} has shown that the Total Variation (TV) loss can be utilized as a regularizer in the loss function to minimize the differences in adjacent pixel values within a patch. Formally, it is defined as follows:
\begin{equation}
    \mathcal{L}_{tv} = \sum_{i, j}\left(\left(p_{i, j}-p_{i+1, j}\right)^2+\left(p_{i, j}-p_{i, j+1}\right)^2\right)^{\frac{1}{2}}
\end{equation}
where $p_{i,j}$ is the pixel values in the patch at position $(i,j)$. This constraint allows the color gradients of the patch to be as smooth as possible. Constraints for other real-world losses will be stated in the next subsection.

\textbf{Defense Evasion.} Other than color discrepancies, we also take potential active defense methods into account: 1) face liveness verification and 2) patch detection, both of which have achieved satisfactory performance by using a pretrained neural network. Correspondingly, we introduce perceptual image patch similarity \cite{zhang2018unreasonable} measured by a pretrained model to serve as a surrogate model of face liveness verification since our inaccessibility to FLV:
\begin{equation}
    \mathcal{L}_{lpips} = \mathbb{E}_{x_i\in D_{c}} \mathrm{LPIPS}(x_i\oplus p,x_v\oplus p)
\end{equation}

To improve FIBA's resistance to patch-based detection, we use Laplacian to measure the inconsistency between the generated patch and the original facial image. Specifically, Laplacian can calculate the second-order spatial derivatives of an image:
\begin{equation}
     \Psi(x) = K \otimes x
\end{equation}
where we define $\otimes$ as convolution operation and $K$ is the Laplacian kernel. Thus, we have the following formulation as a constrain for the edge of the patch:
\begin{equation}
    \mathcal{L}_{edge} = \mathbb{E}_{x_i\in D_{c}} \|\Psi(x)(x_i\oplus p)\|_2
\end{equation}

Till now, the overall loss function can be formalized as:
\begin{equation}
    \mathcal{L} = -\mathcal{L}_{sim} + \alpha\mathcal{L}_{tv} + \beta\mathcal{L}_{edge} + \gamma\mathcal{L}_{lpips}
\end{equation}
where $\alpha$, $\beta$ and $\gamma$ are the coefficients of multi-objective loss and we can minimize $\mathcal{L}$ by optimizing $p$. In this paper, we use Adam \cite{kingma2014adam} optimizer to update the value of $p$. For the selection of $m$, we refer to the result of Figure \ref{fig:mask_region} to use a prefabricated mask for optimization. The performance of different mask regions is given in Section \ref{sec:evaluation}.
 
\subsection{Backdoor Trigger Enhancement}
The key step in FIBA is to generate a backdoor trigger (known as a patch) $p$. An ideal generated patch should also have the following characteristics: 1) high generalization to other models since we cannot get access to FRS under the black-box setting; 2) high robustness to resist distortions made by many factors, e.g., illumination and lens resolution. To achieve this, we utilize several physical enhancement techniques presented in the following subsections.

\textbf{Model Ensemble.} Previous works have demonstrated the effectiveness of model ensemble \cite{chen2023rethinking} in augmenting the generalization of the generated patch within the black-box setting. In this paper, we also adopt this technique while applying our attack to real-world scenarios. Specifically, we modify the $\mathcal{L}_{sim}$ as follows:
\begin{equation}
\begin{aligned}
    \mathcal{L}_{sim} = \mathbb{E}_{x_i\in D_{c}}\frac{1}{M}\sum_{j=1}^M(\mathrm{cos}(\frac{f_{\theta}^{j}(x_i\oplus p)}{\|f_{\theta}^{j}(x_i\oplus p)\|_2},\frac{f_{\theta}^{j}(x_v\oplus p)}{\|f_{\theta}^{j}(x_v\oplus p)\|_2})) 
\end{aligned}
\end{equation}
where $M=\|\mathbb{F}\|$, is the size of assembled models set $\mathbb{F}$ and $f_{\theta}^{j}\in \mathbb{F}$. In practice, we construct our assembled models set including MobileFace \cite{chen2018mobilefacenets}, ResNet50 \cite{he2016deep}, ArcFace \cite{deng2019arcface}, IR50-CosFace \cite{wang2018cosface} and IR50-SphereFace \cite{liu2017sphereface}. All of these models participate in the optimization of the patch and provide the optimizer with more stable gradients to generate a transferable backdoor trigger.

\textbf{Physical Transformation.} As mentioned before, real-world loss should also be considered to mitigate the negative impact of cross-domain transformation. We construct a transformation set $\mathbb{T}$ to simulate this complex process to improve the patch's robustness to digital-to-physical distortions. The first one is ``ColorJiggle"\cite{riba2020kornia} which is used to adjust different brightness or hue changes when applied in real-world scenarios, corresponding to real-world lighting changes. ``RandomAffine" is applied to make the patch translation and rotation invariant, as we can not precisely adjust the disguised facial images to the corresponding position in the physical world. Additionally, motion blur, caused by relative motion between an object or scene and a camera during the exposure time of a photograph or the frame capture of a video, hinders the physical performance of FIBA. Here, we use ``RandomGaussianBlur" to reduce the influence of motion blur and printing errors introduced by the color printer. The pseudocode of the complete algorithm is provided in Appendix \ref{appendix:algorithm}.

\section{EVALUATION}
\label{sec:evaluation}
In this section, experiments are conducted to evaluate the effectiveness of FIBA in the digital (including popular feature extractors and commercial APIs) and the physical domains (IoT devices and commercial APIs). Based on the threat model, results under both white-box and black-box settings are considered.
\subsection{Experimental Setup}
\textbf{Datasets.} Digital experiments of feature extractors focus primarily on datasets: LFW \cite{huang2008labeled}, CALFW \cite{zheng2017cross}, AgeDB \cite{moschoglou2017agedb}, and CelebA-HQ \cite{karras2017progressive} (use CelebA for short). The details of the used datasets are presented in Appendix~\ref{appendix:datasets}.

All of these facial images are resized to 112×112 pixels to match the input requirements of models used for optimization. For experiments in the physical domain, facial images are taken from individuals using edge devices. For the training set of FIBA to generate the backdoor trigger in the real-world evaluation, we use 50 facial images (no overlap with images used for evaluation) from CelebA \textbf{by default} unless otherwise specified and the number of the training data is discussed in Appendix~\ref{appendix:train_size}.

\begin{table*}[t]
\centering
\caption{Overall performance($\times 100\%$) of FIBA against six different face recognition models.}
\label{tab:digital_result}
\begin{threeparttable}
\renewcommand{\arraystretch}{1}
\begin{tabular}{cccccccccc}
\hline
\multirow{3}{*}{\textbf{Source Dataset}} & \multirow{3}{*}{\textbf{Target Model}} & \multicolumn{8}{c}{\textbf{Target Dataset}} \\ \cline{3-10} 
 &  & \multicolumn{2}{c}{CelebA-HQ} & \multicolumn{2}{c}{LFW} & \multicolumn{2}{c}{AgeDB} & \multicolumn{2}{c}{CALFW} \\ \cline{3-10} 
 &  & w/o attack & w/ attack & w/o attack & w/ attack & w/o attack & w/ attack & w/o attack & w/ attack \\ \hline
\multirow{6}{*}{CelebA-HQ} & $\text{MobileFace}^{*}$ & 0.14 & 100.00 & 0.00 & 100.00 & 0.00 & 100.00 & 0.00 & 100.00 \\
 & MobilenetV2 & 0.18 & 98.10 & 0.00 & 92.23 & 0.00 & 95.08 & 0.00 & 96.10 \\
 & ShuffleNetV1 & 0.17 & 99.96 & 0.00 & 98.57 & 0.00 & 99.98 & 0.02 & 97.78 \\
 & IR50-Softmax & 0.49 & 99.63 & 0.14 & 98.53 & 0.00 & 99.27 & 0.16 & 98.41 \\
 & IR50-SphereFace & 0.69 & 99.75 & 0.31 & 98.75 & 0.55 & 99.24 & 0.65 & 97.90 \\
 & CASIA-Softmax & 0.82 & 99.94 & 0.27 & 99.67 & 0.20 & 99.84 & 0.55 & 98.84 \\ \hline
\multirow{6}{*}{LFW} 
& $\text{MobileFace}^{*}$ & 0.02 & 100.00 & 0.02 & 100.00 & 0.00 & 100.00 & 0.02 & 100.00 \\
 & MobilenetV2 & 0.00 & 69.06 & 0.33 & 73.59 & 0.00 & 65.90 & 0.04 & 86.31 \\
 & ShuffleNetV1 & 0.00 & 72.27 & 0.02 & 78.73 & 0.00 & 70.14 & 0.04 & 54.20 \\
 & IR50-Softmax & 0.29 & 76.41 & 0.67 & 82.80 & 0.08 & 71.90 & 0.55 & 68.20 \\
 & IR50-SphereFace & 0.29 & 82.14 & 0.53 & 81.55 & 0.16 & 78.61 & 0.71 & 51.86 \\
 & CASIA-Softmax & 0.31 & 77.75 & 1.41 & 84.16 & 0.24 & 74.12 & 1.53 & 71.75 \\ \hline
\multirow{6}{*}{AgeDB} 
& $\text{MobileFace}^{*}$ & 0.00 & 100.00 & 0.00 & 100.00 & 0.02 & 100.00 & 0.00 & 100.00 \\
 & MobilenetV2 & 0.00 & 94.57 & 0.00 & 81.90 & 0.02 & 99.75 & 0.00 & 85.96 \\
 & ShuffleNetV1 & 0.00 & 96.45 & 0.00 & 87.80 & 0.02 & 99.96 & 0.00 & 96.67 \\
 & IR50-Softmax & 0.06 & 92.98 & 0.10 & 86.33 & 0.47 & 97.22 & 0.14 & 89.96 \\
 & IR50-SphereFace & 0.27 & 96.14 & 0.45 & 85.92 & 1.18 & 99.53 & 0.59 & 95.37 \\
 & CASIA-Softmax & 0.45 & 97.94 & 0.51 & 94.49 & 1.18 & 99.98 & 0.51 & 95.55 \\ \hline
\multirow{6}{*}{CALFW} 
& $\text{MobileFace}^{*}$ & 0.00 & 100.00 & 0.00 & 100.00 & 0.00 & 100.00 & 0.00 & 100.00 \\
 & MobilenetV2 & 0.00 & 84.63 & 0.20 & 86.71 & 0.00 & 73.47 & 0.07 & 63.06 \\
 & ShuffleNetV1 & 0.00 & 87.04 & 0.00 & 87.41 & 0.00 & 82.78 & 0.00 & 73.86 \\
 & IR50-Softmax & 0.04 & 80.61 & 0.29 & 84.10 & 0.02 & 74.53 & 0.16 & 83.65 \\
 & IR50-SphereFace & 0.33 & 83.53 & 0.73 & 87.61 & 0.25 & 80.39 & 0.51 & 88.16 \\
 & CASIA-Softmax & 0.16 & 91.22 & 1.20 & 95.57 & 0.20 & 87.57 & 1.10 & 87.69 \\ \hline
\end{tabular}
\begin{tablenotes}
\footnotesize
\item [1] Models with $*$ denote the white-box attack (source model is the same as the target model).
\end{tablenotes}
\end{threeparttable}
\end{table*}

\textbf{Target Models, APIs and IoT Devices.}
We use six common facial feature extractors in our evaluation, with detailed information presented in Table \ref{tab:model_info}. Since the thresholds of these models are different, we unify the threshold ($\mathrm{\delta_{match}}=0.35$) for simplicity, and a higher threshold also presents a more conservative attack performance and lower false positive rate (evaluation with higher $\mathrm{\delta_{match}}$ can be found in Figure \ref{fig:threshold_performance}). To make the evaluation more practical, we also use five commercial APIs from different vendors: Aliyun (version: 20191230), Face++ (version: v4), ArcSoft (version: v3), Huawei (version: v2), and Tencent (version: v3). These vendors are selected since they are widely deployed and have become the mainstream of face recognition services. All of these APIs claim to achieve over 99\% accuracy with less than 0.1\% false positive rate. Edge devices are also considered since they are a common part of FRS. To simplify the expression, IoT devices from SenseTime, HIKVISION, and MoreDian are labeled as D-1, D-2, and D-3 for short, respectively. More details can be found in Table \ref{tab:device_config}.

\textbf{Baseline Attack.} Will FIBA outperform traditional adversarial attacks? To answer this question, we conduct experiments to compare FIBA with universal adversarial attacks. A key difference between FIBA and the adversarial attacks used in our paper is that we modify the multi-objective similarity loss as follows:
\begin{equation}
\begin{aligned}
    \mathcal{L}_{sim} = \mathbb{E}_{x_i\in D_{c}}\mathrm{cos}(\frac{f_{\theta}(x\oplus p)}{\|f_{\theta}(x\oplus p)\|_2},\frac{f_{\theta}(x_v)}{\|f_{\theta}(x_v)\|_2})
\end{aligned}
\end{equation}
This loss function aims to maximize the similarity between disguised attacker $x_i\oplus p$ and insider $x_v$. The generated patch aims to make any attackers who are disguised with $p$ close to the enrolled insider's facial image $x_v$. Other settings are the same as those for FIBA.

\textbf{Evaluation Metrics.} To measure the effectiveness of FIBA with a test set $D_b$, we define ASR as:
\begin{equation}
\begin{aligned}
    ASR = \frac{1}{N}\sum_{i=0}^{N} \mathbb{I}\left(\cos \left(f_{\theta}(x_i\oplus p), f_{\theta}(x_v\oplus p)\right) \geq \mathrm{\delta_{match}}\right)
\end{aligned}
\end{equation}
where $\mathbb{I}(\cdot)$ is the indicator function. Other factors, such as face detection success rate, are also given in the experiments.

\subsection{Attack Performance}
\textbf{White-box and Black-box Attack in Digital Domain.} In this subsection, for each dataset, patches are generated for 5 different identities (insiders) and tested with 500 facial images from other identities that differed from the training set of patch generation. It is noteworthy that MobileFace is used as our surrogate model since it's widely used for face recognition tasks with high accuracy in face verification while ensuring the model is small in size and fast in execution. Other models act as black-box models to test the transferability of FIBA, with 50 facial images randomly selected for patch training during each iteration. We also evaluate the ASR of FIBA when the source dataset (training set) and the target dataset (evaluation set) are not the same since face images from different datasets have a distribution shift that can negatively impact the patch's optimization.

The results are presented in Table \ref{tab:digital_result}. For comparison, the match rate of the benign faces is also given. Note that FIBA can achieve an ASR of 100\% under a white-box setting even if the target dataset (first row for different source datasets) is inaccessible. Under the black-box setting, where the victim model is inaccessible, FIBA can still achieve remarkable performance with more than 99\% for most combinations if CelebA is selected as the source dataset. When other datasets are chosen as the source dataset to train a backdoor trigger, there is a slight decrease in ASR. The observed decrease in ASR can be attributed to the fact that the other three datasets have a lower quality than CelebA, and only a small portion of these datasets is included in our training set. Consequently, the quality of the training set is pivotal to the optimization process of FIBA, and adversaries may feasibly gather such images (for instance, from CelebA) through artificial means. It's also noteworthy that different models present different vulnerabilities under the black-box setting. $\text{CASIA-Softmax}$ exhibits greater susceptibility, whereas $\text{MobilenetV2}$ demonstrates improved robustness against FIBA. Overall, FIBA proves to be effective and feasible for attackers, as no additional computation is needed once the backdoor trigger has been generated. The cross-model and cross-dataset results further demonstrate FIBA's high generalization and universality in the digital domain. 

\begin{table}[t]
\centering
\caption{ASR($\times 100\%$) of adversarial attack.}
\label{tab:baseline_attack}
\renewcommand{\arraystretch}{1.1}
\begin{tabular}{cclcc}
\hline
\multirow{2}{*}{\textbf{Target Model}} & \multicolumn{4}{c}{\textbf{Target Dataset}} \\ \cline{2-5} 
 & \multicolumn{1}{l}{CelebA-HQ} & \multicolumn{1}{c}{LFW} & AgeDB & \multicolumn{1}{l}{CALFW} \\ \hline
MobileFace & 96.76 & 93.53 & 87.67 & 81.76 \\
MobilenetV2 & 23.55 & 23.53 & 9.72 & 0.78 \\
ShuffleNetV1 & 32.80 & 17.84 & 10.20 & 0.20 \\
IR50-Softmax & 63.18 & 43.33 & 50.50 & 11.37 \\
IR50-SphereFace & 67.00 & 71.18 & 51.48 & 13.53 \\
CASIA-Softmax & 68.88 & 85.49 & 50.73 & 13.73 \\ \hline
\end{tabular}
\end{table}

\begin{figure}[t]
  \centering
  \includegraphics[width=0.9\linewidth]{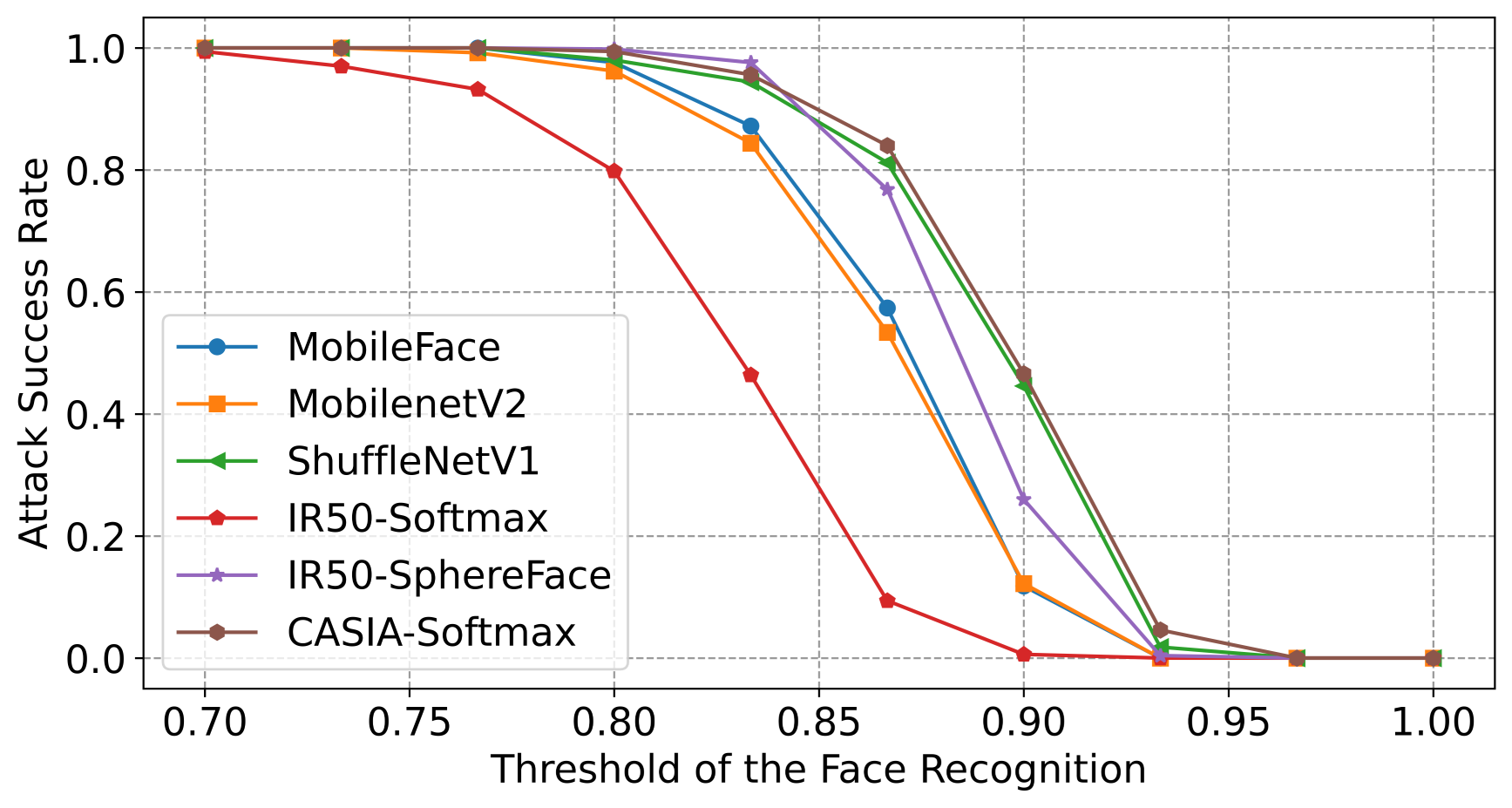}
  \caption{Impact of thresholds on ASR (surrogate model: MobielFace, source dataset: CelebA).}
  \label{fig:threshold_performance}
\end{figure}

\textbf{Comparison with Adversarial Attack.}
The performance of the adversarial attacks for comparison is also presented in Table \ref{tab:baseline_attack}. MobileFace and CelebA serve as surrogate models and source datasets to generate adversarial patches. Under the white-box setting, the baseline attack demonstrates strong performance, achieving an ASR of 96.76\%, 93.53\%, 87.67\%, and 81.76\%, respectively, for four datasets. However, when the information on the target model is unavailable, the ASR declines significantly. For the CALFW dataset on ShuffleNetV1, the adversarial attack achieves a mere ASR of 0.2\%, compared to FIBA's 97.78\%.

As shown in Table \ref{tab:baseline_attack}, there is a significant performance gap exists between FIBA and this adversarial attack. The disparity can be attributed to FIBA's ability to exploit intrinsic flaws in FRS to substitute key facial features with generated trigger features. In contrast, adversarial patch-based attacks or previous enrollment-stage attacks \cite{wuuniid,li2020light} have to balance the similarity trade-off of different identities. To make it clear, we visualize the facial features of the benign, triggered, and enrolled sample in Figure \ref{fig:tsne}. Upon applying dimensionality reduction, it becomes evident that as the benign face image is masked with a backdoor trigger, its features intuitively converge with those of the enrolled sample. Additionally, we have discussed how different sizes of training data influence the performance of FIBA (details shown in Appendix \ref{appendix:train_size}) and found out that even one image is enough to train a backdoor trigger with more than 90\% ASR within a black-box setting. This also distinguishes FIBA from universal adversarial attacks.

\begin{table}[t]
\centering
\caption{Face detection success rate($\times 100\%$) of FIBA.}
\label{tab:face_detection}
\renewcommand{\arraystretch}{1.1}
\begin{tabular}{ccccc}
\toprule
\multirow{2}{*}{\textbf{Face Detection}} & \multicolumn{4}{c}{\textbf{Train Dataset}} \\ \cline{2-5} 
 & CelebA-HQ & LFW & AgeDB & CALFW \\ \hline
MTCNN & 99.00 & 91.80 & 97.60 & 94.80 \\
Baidu API & 100.00 & 99.00 & 99.00 & 99.00 \\
Alibaba API & 94.00 & 90.00 & 80.00 & 86.00 \\
\bottomrule
\end{tabular}
\end{table}

\begin{figure}[t]
    \centering
    \includegraphics[width=0.9\linewidth]{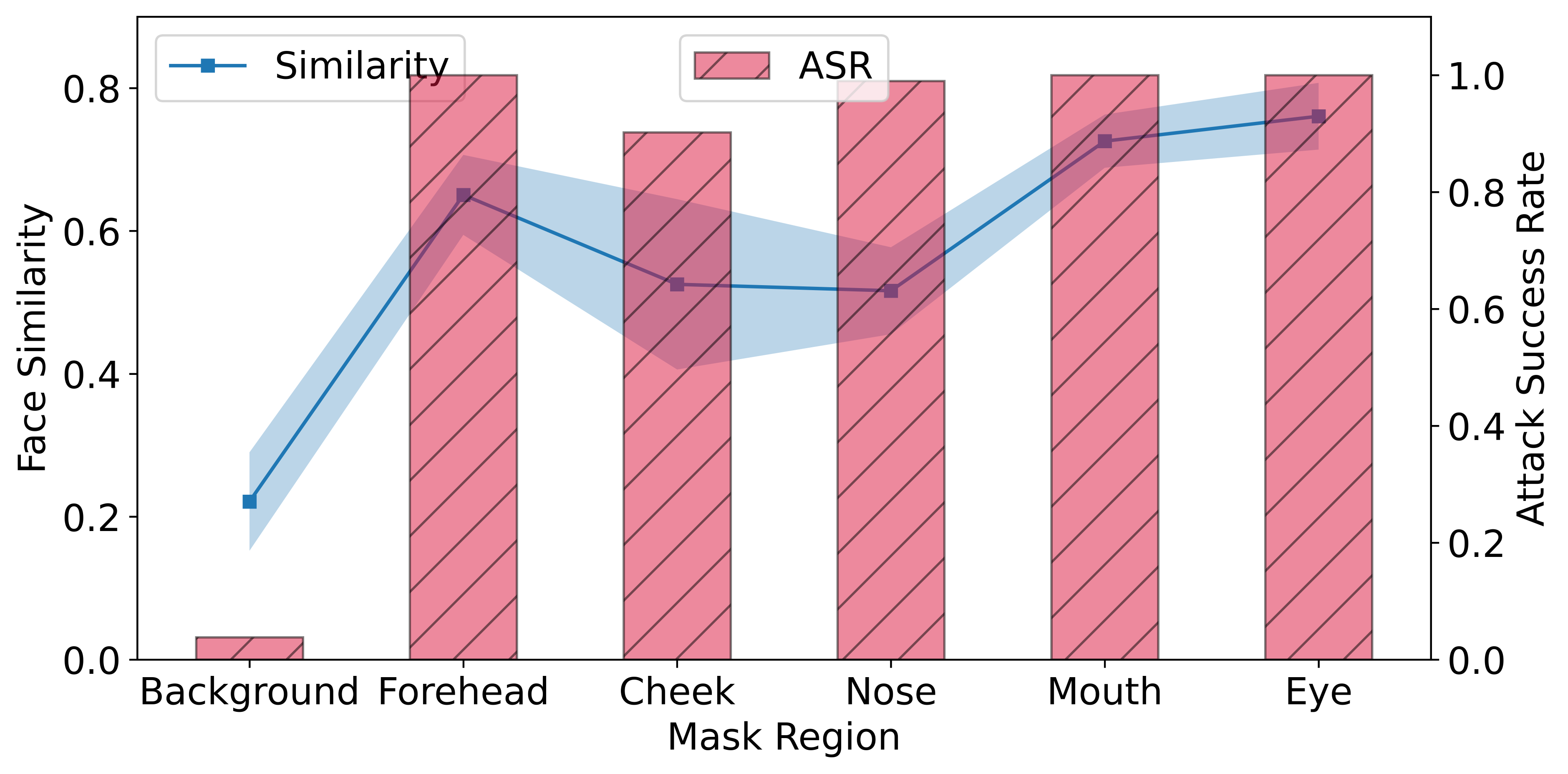}
    \caption{Attack performance using mask from different facial regions.}
    \label{fig:mask_performance}
\end{figure}

\textbf{Attack Performance with Different Threshold.} In the previous evaluation, we set $\mathrm{\delta_{match}}$ to 0.35. To comprehend this factor, we analyzed how $\mathrm{\delta_{match}}$ affects the attack performance with the experimental results on CelebA (target dataset) shown in Figure \ref{fig:threshold_performance}. We can observe from the figure that most of the similarity under the black-box setting can achieve 0.7 since the ASR of most models can approach 100\%. Consequently, FIBA can maintain a high ASR even when the threshold approaches 0.7, which also signifies the high transferability of FIBA across various models.

\textbf{Face Detection Rate of FIBA.} Given that FIBA involves the disguise of the insider or attackers using a backdoor trigger, it is crucial to explore whether the disguised face can navigate through the key step of FRS: face detection. Similarly, we use MobileFace and CelebA to generate a backdoor trigger and test the face detection rate (FDR) on a pretrained detection model MTCNN \cite{zhang2016joint} and two commercial APIs. From Table \ref{tab:face_detection}, we can observe that FIBA can achieve high FDR, the first step to attack FRS.

\textbf{Attack Performance of Different Mask Regions.} Note that though a specific mask has been selected to optimize the trigger in the previous section, how this factor impacts the performance of FIBA is unknown. Therefore, to better understand how the choice of different mask regions influences the performance of FIBA, we conducted experiments using MobileFace and CelebA (50 and 500 samples for training and evaluation). Results displayed in Figure \ref{fig:mask_performance} illustrate that mask's region has varying impact on ASR and average similarity. Specifically, the mask of the eye's region achieves the highest ASR, and the background region has the most minor improvement on ASR. By replacing key features of the eye region with a backdoor trigger, the average similarity can achieve more than 0.8. Thus, the mask of the eye region is the most suitable area to optimize the trigger.

\begin{table}[t]
\centering
\caption{Digital attack performance on commercial APIs.}
\label{tab:api_performance}
\renewcommand{\arraystretch}{1.1}
\begin{tabular}{cccc}
\hline
\textbf{API Vendors} & $\mathrm{\delta_{match}}$ & Similarity & ASR ($\times 100\%$) \\ \hline
Huawei & 0.93 & 0.90 & 18.55 \\
ArcSoft & 0.80 & 0.98 & 100.00 \\
Alibaba & 0.61 & 0.75 & 99.60 \\
Face++ &  0.62 &  0.89 & 100.00 \\
Tencent & 0.40 & 0.40 & 42.66 \\ \hline
\end{tabular}
\end{table}

\begin{figure}[t]
    \centering
    \includegraphics[width=0.9\linewidth]{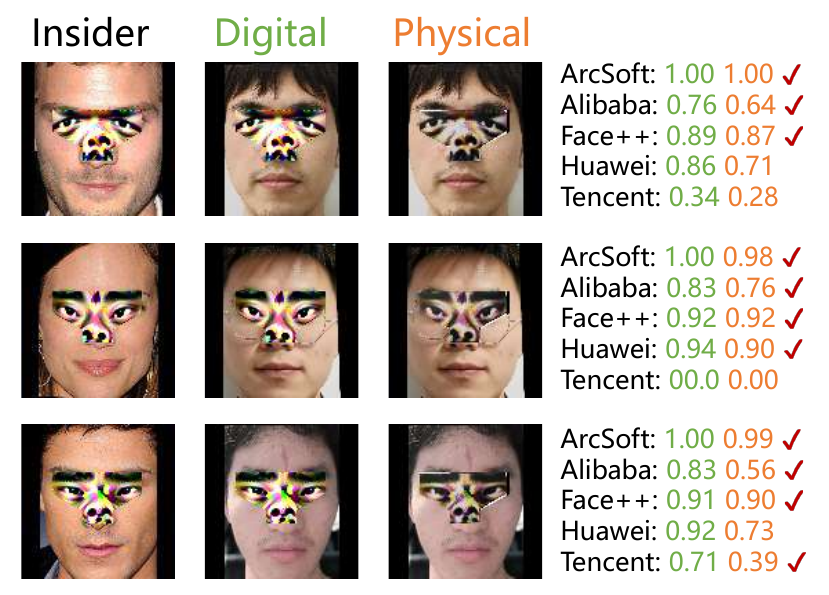}
    \caption{Examples of FIBA domain against five APIs. \ding{52} indicates that the similarity score exceeds the threshold in either the digital or physical domain.}
    \label{fig:demo}
\end{figure}

 \subsection{Real-world Evaluation}
In this subsection, we endeavor to evaluate the effectiveness of FIBA and assess the robustness of the commercial APIs or IoT devices in real-world scenarios.
 
\textbf{API Evaluation in Digital Domain.} We employ the model ensemble technique as mentioned in Section \ref{sec:method} to generate a backdoor trigger (50 images for training) and evaluate its efficacy with 500 samples from CelebA (different identities than those in the training set). The results are presented in Table \ref{tab:api_performance}. Face similarity thresholds of different vendors' FRS services and the average similarity of FIBA are also given in the table. In scenarios where the API returns ``no face detected" for a given sample, said sample is categorized as a failed sample. An attack is deemed successful solely if the response similarity exceeds the given threshold of the corresponding vendor. Note that for APIs from three vendors: ArcSoft, Alibaba, and Face++, FIBA is able to achieve nearly 100\% ASR. Additionally, the similarity is much higher than $\mathrm{\delta_{match}}$. However, for APIs from Huawei and Tencent, FIBA exhibits inferior attack performance. This could be attributed to these vendors having utilized certain detection methods or deployed robust models. Illustrations (not included in evaluation) of FIBA are depicted in Figure \ref{fig:demo}.

\begin{table}[t]
\centering
\caption{Physical authentication attack performance on commercial APIs.}
\label{tab:api_performance_physical}
\renewcommand{\arraystretch}{1.1}
\begin{tabular}{cccc}
\hline
\textbf{Enrollment} & \textbf{Vendors} & Similarity & ASR ($\times 100\%$) \\ 
\hline
\multirow{5}{*}{Digital} & Huawei & 0.78 & 16.00 (4/25) \\
& ArcSoft & 0.94 & 96.00 (24/25) \\
& Alibaba & 0.75 & 60.00 (15/25) \\
& Face++ & 0.84 & 92.00 (23/25) \\
& Tencent & 0.38 & 24.00 (6/25) \\ \hline
\multirow{5}{*}{Physical} & Huawei & 0.91 & 8.00 (2/25) \\
& ArcSoft & 0.93 & 96.00 (24/25) \\
& Alibaba & 0.75 & 56.00 (14/25) \\
& Face++ & 0.89 & 96.00 (24/25) \\
& Tencent & 0.38 & 32.00 (8/25) \\ \hline
\end{tabular}
\end{table}

\textbf{API Evaluation in the Physical Domain.} As previously mentioned in Section \ref{sec:threat}, the threat model of FIBA in the physical domain is categorized into two primary approaches: 1) digital enrollment and physical authentication; 2) physical enrollment and physical authentication. Consequently, we enlisted 5 volunteers and obtained their facial images to create backdoor triggers for each of them. Another 5 volunteers wearing printed triggers participated in face authentication in the physical domain. In total, we have 25 pairs forming a test set to evaluate FIBA in the physical domain. For the digital and physical enrollment experiments, we enroll the disguised facial images from the digital domain or taken in the physical domain into the database.
 
The results are shown in Table \ref{tab:api_performance_physical}. In contrast to digital authentication, both the ASR and the average similarity of FIBA have experienced declines to varying degrees, notably for the API from ``Alibaba". However, FIBA can still attack ``ArcSoft" and ``Face++" with more than 90\% ASR, and it's unbearable for some safety-critical applications that have been equipped with these APIs. Also, there hasn't been a big drop in ASR between the digital and physical domains, with even a slight increase in average similarity for several APIs. We suppose that's because distortions between the physical and digital backdoor trigger counteracts the negative effect raised by physical authentication, which also validates the robustness of FIBA against physical distortion.

\begin{figure}[t]
    \centering
    \includegraphics[width=0.9\linewidth]{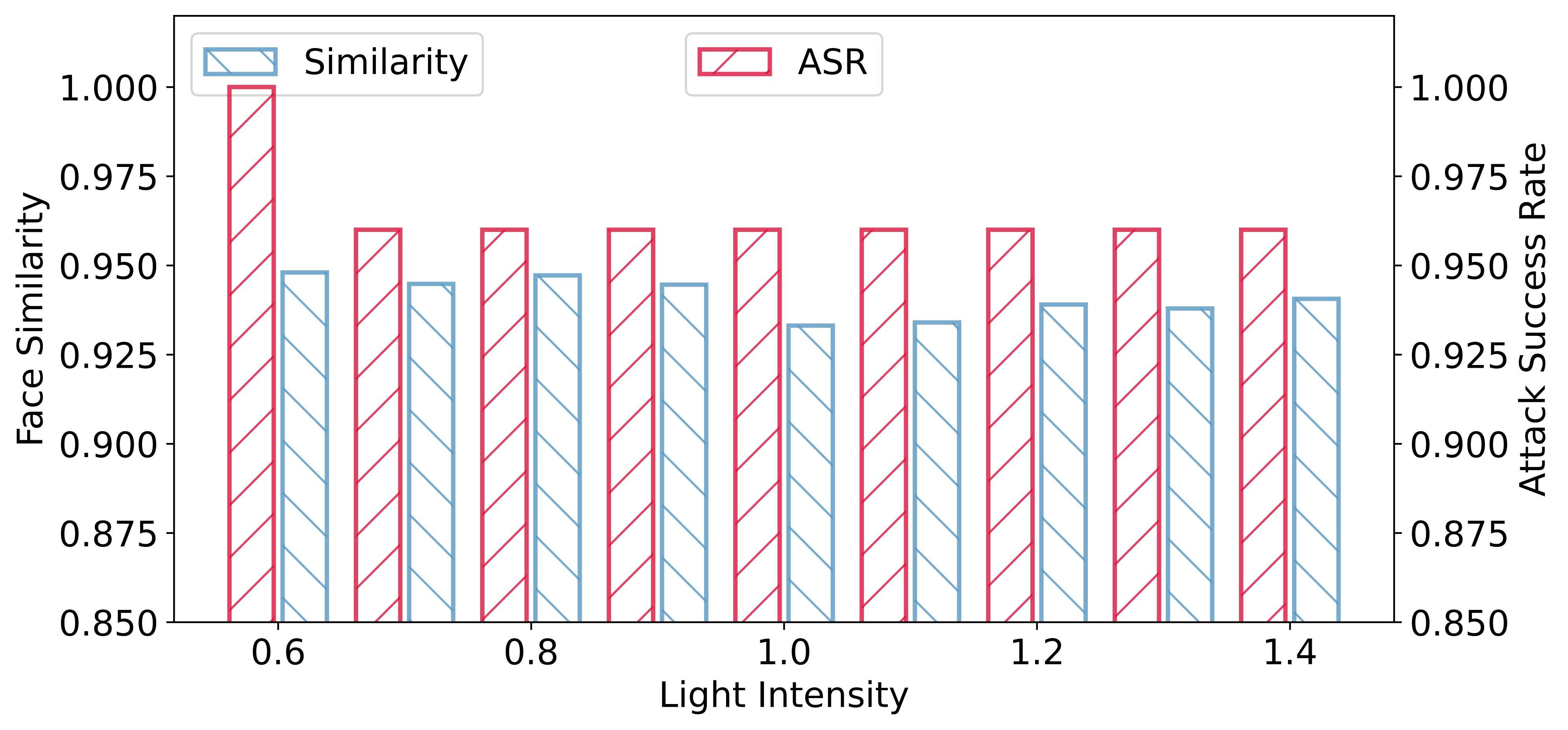}
    \caption{Impact of light conditions on FIBA.}
    \label{fig:light_physical}
\end{figure}

\textbf{Impact of Light Condition.}
Another important factor that needs to be studied in the physical domain is that the environment's lighting conditions cannot be guaranteed and pose a significant challenge to existing attacks. To evaluate the impact of different lighting conditions on FIBA, we conducted experiments with varying lighting conditions, and all other settings are the same as those in the physical domain (physical enrollment and authentication with the ArcSoft API). More details are presented in Appendix \ref{appendix:physical}.

The experimental results presented in Figure \ref{fig:light_physical} illustrate that FIBA can maintain an ASR of more than 95\% under varying lighting conditions, and the average face similarity remains stable too. Note that on the horizontal axis, "1" denotes a light intensity equivalent to the normal indoor lighting intensity. Even under extremely dark or overly bright conditions, the ASR remains unchanged compared to that achieved under normal lighting conditions. Through these experiments, we confirm the attack performance of FIBA under different lighting conditions, demonstrating that FIBA is resilient to lighting variations in the real world.

\begin{figure}[t]
    \centering
    \includegraphics[width=0.9\linewidth]{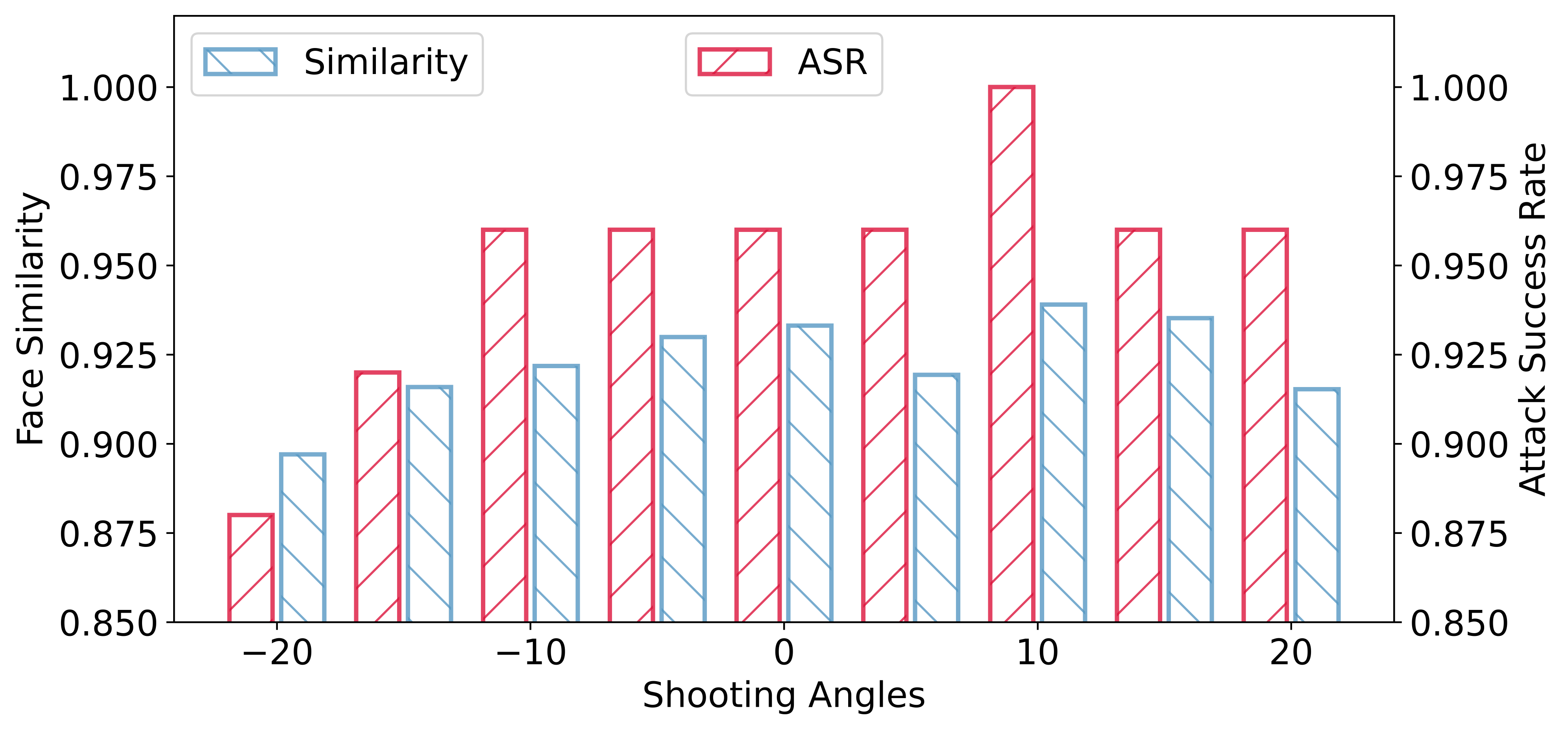}
    \caption{Impact of shooting angles on FIBA.}
    \label{fig:angles_physical}
\end{figure}

\textbf{Impact of Shooting Angles.}
Apart from environmental factors such as lighting conditions, shooting angles also play a significant role during the authentication stage. Attacks that can only perform well with specific shooting angles are not applicable in the real world. To evaluate the performance of FIBA against different shooting angles in the physical domain, we conduct an experiment using ArcSoft API with the same settings as in the previous evaluation. More details are presented in Appendix \ref{appendix:physical}. 

Results are presented in Figure \ref{fig:angles_physical}, where the horizontal axis denotes different shooting angles (ranging from -20 to 20 degrees), and a value of 0 indicates the photographer is facing the camera directly. Intuitively, with different shooting angles, the ASR of FIBA also fluctuates and reaches a peak when the shooting angle is at 10 degrees. Considering it as a whole, the ASR of FIBA has consistently remained above 85\%, demonstrating its stability and insensitivity to different spatial locations.

\subsection{Evaluation on IoT Devices}
In addition to API tests, IoT devices are included in our experiments due to their advanced components (e.g., infrared-based face liveness detection, robustness-enhanced model) that contribute to a more reliable FRS. These features pose a significant challenge to the existing attack methods. Therefore, to further assess the performance of FIBA in complex real-world settings, we conducted attack experiments on three IoT devices. Detailed information about these three IoT devices can be found in Table \ref{tab:device_config}, and all of them feature a deployed face liveness verification algorithm. Furthermore, these three devices offer users varied methods for enrollment, such as taking and uploading photos. For the first method, we photographed a disguised volunteer and registered their face on the device. For the second method, we directly enroll a digital image of the triggered face on the device. Before authentication, we simulated the enrollment of insiders by registering five disguised faces on each device respectively. Later, the triggers of the insiders are distributed to another five volunteers (different identities with insiders) recruited to act as unknown attackers. It should be noted that during authentication, attackers are allowed to adjust the location of their backdoor triggers for launching attacks for at most 3 times; when the attacker wearing the trigger $p_t$ is recognized as the corresponding insider $x_k$, and $k=t$, then the attack is labeled a success. Moreover, we have noticed that these devices claim to support face recognition while wearing face masks. Therefore, we also evaluate the performance of FIBA when the attacker is wearing a face mask and a backdoor trigger simultaneously.

The attack results depicted in Table \ref{tab:device_performance} illustrate the remarkable performance in spoofing IoT devices equipped with FLV. For attacks conducted without a face mask, FIBA has achieved ASR of 42\%, 100\%, and 20\% on the three devices, respectively, which suggests that if FIBA is really to be launched by attackers in the real world, these edge devices could be circumvented by disguised attackers with a high probability. Notably, a 100\% ASR is achieved on all IoT devices if attackers launch FIBA wearing face masks and triggers simultaneously during authentication. This is because the remaining distinguished features of the original face are obscured by the face mask, leaving only the features of the backdoor trigger on which FRS can base its decision. This phenomenon indicates that, although many face recognition service vendors have claimed that their devices can maintain high accuracy with a low false positive rate when wearing a face mask, this improvement \textbf{exposes the vulnerabilities of FRS to potential attacks such as FIBA!}

\begin{table}[t]
\centering
\caption{Performance of FIBA on IoT devices.}
\label{tab:device_performance}
\small
\renewcommand{\arraystretch}{1.1}
\begin{tabular}{ccc}
\hline
\textbf{ID} & ASR($\times 100\%$) w/o mask  & ASR($\times 100\%$) w/ mask \\ \hline
D-1 &  42.00(13/25) & 100.00 (25/25) \\
D-2 & 100.00(25/25) & 100.00 (25/25) \\
D-3 & 20.00(5/25) & 100.00 (25/25) \\ 
\hline
\end{tabular}
\end{table}

\section{IMPLICATION}
\label{sec:implication}
In this section, based on FIBA's real-world evaluation, we offer several suggestions for service vendors. Subsequently, we discuss FIBA's countermeasures and limitations.
\subsection{Suggestions for Service Vendors}
Our analysis has pinpointed vulnerabilities within certain commercial APIs and IoT devices. Moreover, investigating methods to bolster the existing FRS through model refinement and attack detection is paramount in effectively addressing these vulnerabilities. Specifically, distinct methodologies should be employed at the enrollment and authentication stages to address their challenges and vulnerabilities.

\textbf{Suggestions of Enrollment Stage.} (1) Data quality assurance: High-quality data must be ensured, as low-quality enrolled images can result in an elevated false positive rate during recognition. (2) Face liveness detection: Given the rapid advancements in generative models \cite{rombach2022high}, enrolled faces can potentially be forged using sophisticated algorithms. For real-time face registration, employing dynamic face liveness detection is necessary. (3) Patch detection: Considering attack costs, many attackers opt for patch-based attacks as a means of disguise. Implementing patch detection algorithms is critical in thwarting these attacks, and physical distortion should also be considered during detection.

\textbf{Suggestions of Authentication Stage.} (1) Multi-factor Authentication (MFA): It is advisable to combine facial recognition with another authentication form (such as PINs or biometric signals) whenever possible, to bolster security. (2) Anomaly Detection Systems: Deploying systems capable of identifying unusual authentication attempts (such as patch attacks) or patterns indicative of a potential security breach can enable proactive security measures. (3) Active Prompt: If the user is wearing a face mask, it is better to prompt the user to remove the mask; otherwise, this will bring more false positives for recognition and significantly increase the ASR of attackers.

We have reported our experimental findings and suggestions to affected service vendors. We hope our work can help service vendors strengthen FRS's robustness and they must also consider whether these challenges extend beyond FRS to other biometric recognition technologies, necessitating a broader examination of the reliability and security of such systems. Addressing these profound questions is imperative to enhance trust in biometric authentication and ultimately pave the way for seamless and trustworthy identification.

\subsection{Potential Defenses}
Note that FIBA differs from traditional backdoor attacks and it exploits the natural flaws of FRS to launch attacks. Therefore, traditional backdoor defense methods~\cite{wang2019neural,gao2019strip,liu2019abs} are not applicable to FIBA. Given the scarcity of defense methods for attacks such as FIBA, we undertake a preliminary exploration of potential defenses to mitigate FIBA. We focus on enhancing the models' inherent robustness instead of detection-based defenses. 

\textbf{Adversarial Training.} Different from traditional adversarial training, we adjust its objective and devise a min-max optimization algorithm to defend FIBA. Since FIBA exploits the facial feature extractor's characteristic that the extractor can no longer separate faces by adding perturbations, we simulate the generation of perturbations and force the extractor to distinguish them. For a batch of facial images $x_b$, we can maximize their pair-wise similarity by adding a universal small perturbation $z$:
\begin{equation}
    \mathcal{L}_{ps}(x_b,z)= \|(f_{\theta}(x_b+z)\cdot f_{\theta}(x_b+z)^{T})\odot(\mathrm{1}-\mathbf{E})\|
\end{equation}
where $\mathbf{E}$ is an identity matrix. The process of perturbation generation can be formalized as:
\begin{equation}
\begin{aligned}
    \delta_z &=\nabla_z \mathcal{L}_{ps}(x_b,z)
\end{aligned}
\end{equation}
Therefore, $z$ can be update with  $z + \alpha\cdot\frac{\delta_z}{\|\delta_z\|_2}$. We can optimize the perturbation by maximizing the pair-wise similarity within a batch. Then we try to force $f_{\theta}$ to distinguish facial features adversarially:
\begin{equation}
\begin{aligned}
    \arg\min\limits_{\theta}\ &\mathrm{CE}(h(f_{\theta}(x_b+z)),y_b)+\mathcal{L}_{ps}(x_b,z)
\end{aligned}
\end{equation}
where $h(\cdot)$ is a linear layer to map embedding to the number of classes and $\mathrm{CE}(\cdot)$ is the cross entropy loss. The complete algorithm can be found in Appendix~\ref{appendix:defense}. For evaluation, we finetune the pretrained IR50-Softmax model with the above loss function. In order to evaluate defense efficacy against FIBA, we generate a backdoor trigger in advance with MobileFace on CelebA, utilizing 50 images for training, and transfer the attack to a robustness-enhanced model. The performance of the defense strategy is measured by the average similarity of the triggered samples. 

\begin{figure}[t]
    \centering
    \includegraphics[width=0.9\linewidth]{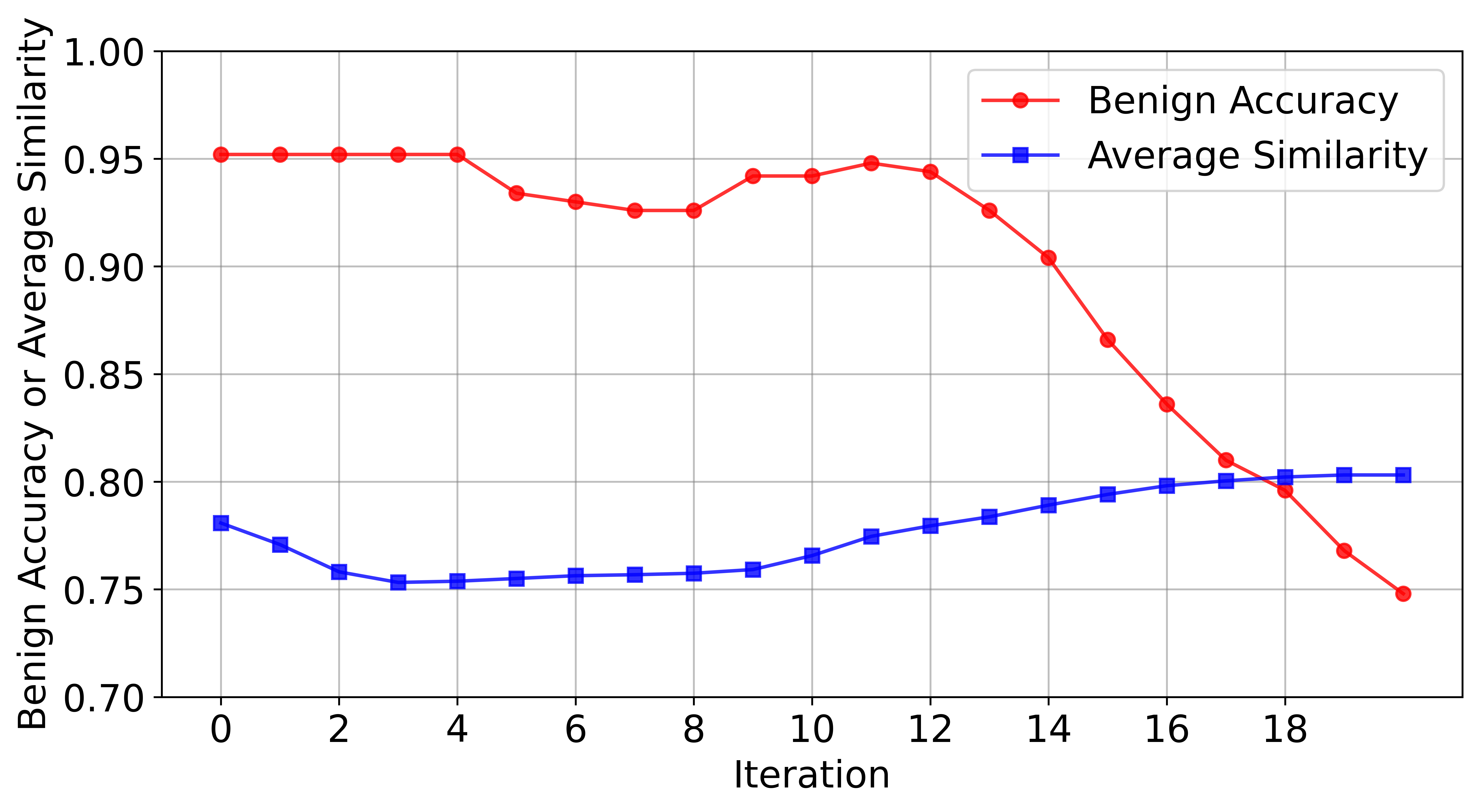}
    \caption{Defense performance with different iteration.}
    \label{fig:defense_performance}
\end{figure}

The performance with benign accuracy and average attack similarity of this scheme during the finetuning process is shown in Figure~\ref{fig:defense_performance}. We can see from the figure that the average similarity does decrease at the beginning. However, as this process continues, benign accuracy exhibits a slight decrease as well, which also leads to the increase of the average similarity correspondingly. This interacting process ultimately results in a rapid decline of benign accuracy. The dynamic change of benign accuracy and average similarity demonstrates that it's hard to achieve both high accuracy and defense performance at the same time. We suppose that since traditional facial feature extractor training forces the extractor to distinguish different identities, the extractor has overfitted by focusing on some local parts of different faces to achieve high accuracy. However, the defense strategy disrupts the overfitting, with a decrease in benign accuracy. The extractor that lost the ability to extract distinguished features from facial images will also fail to defend FIBA.

\subsection{Limitations}
The limitations of our attack are concluded as follows. (1) FIBA solely relies on a 2D printer to produce the trigger, reducing attack performance and increasing the risk of FLV detection. To tackle this problem, we can optimize the trigger in 3D dimensions and print it with a 3D printer to make it more stealthy. (2) The transferability of FIBA is limited by the capabilities of the surrogate models. This can also be alleviated by combining some advanced techniques \cite{zhu2022toward} to improve its transferability.

\section{CONCLUSION}
In this paper, we reveal the natural flaws of face recognition systems based on user study and experimental results. With this observation, we propose FIBA, an enroll-stage facial identity backdoor attack. Once a compromised insider of FRS has enrolled his disguised face (wearing a backdoor trigger) into the facial feature database, any attackers wearing a backdoor trigger can bypass the FRS. Experiments on six models, five APIs, and three IoT devices across four datasets validate the remarkable performance of FIBA in digital and physical domains. Our future direction is to devise mitigation against FIBA and construct a reliable and trustworthy FRS.

\newpage

\bibliography{main}

\begin{thebibliography}{10}
\providecommand{\url}[1]{#1}
\csname url@samestyle\endcsname
\providecommand{\newblock}{\relax}
\providecommand{\bibinfo}[2]{#2}
\providecommand{\BIBentrySTDinterwordspacing}{\spaceskip=0pt\relax}
\providecommand{\BIBentryALTinterwordstretchfactor}{4}
\providecommand{\BIBentryALTinterwordspacing}{\spaceskip=\fontdimen2\font plus
\BIBentryALTinterwordstretchfactor\fontdimen3\font minus \fontdimen4\font\relax}
\providecommand{\BIBforeignlanguage}[2]{{%
\expandafter\ifx\csname l@#1\endcsname\relax
\typeout{** WARNING: IEEEtran.bst: No hyphenation pattern has been}%
\typeout{** loaded for the language `#1'. Using the pattern for}%
\typeout{** the default language instead.}%
\else
\language=\csname l@#1\endcsname
\fi
#2}}
\providecommand{\BIBdecl}{\relax}
\BIBdecl

\bibitem{zhao2003face}
W.~Zhao, R.~Chellappa, P.~J. Phillips, and A.~Rosenfeld, ``Face recognition: A literature survey,'' \emph{ACM Computing Surveys}, vol.~35, no.~4, pp. 399--458, 2003.

\bibitem{kamgar2011toward}
B.~Kamgar-Parsi, W.~Lawson, and B.~Kamgar-Parsi, ``Toward development of a face recognition system for watchlist surveillance,'' \emph{IEEE Transactions on Pattern Analysis and Machine Intelligence}, vol.~33, no.~10, pp. 1925--1937, 2011.

\bibitem{32FRStatistics}
M.~Calvello, ``32 facial recognition statistics to know in 2023,'' \url{https://www.g2.com/articles/facial-recognition-statistics}.

\bibitem{wuuniid}
Z.~Wu, Y.~Cheng, S.~Zhang, X.~Ji, and W.~Xu, ``Uniid: Spoofing face authentication system by universal identity,'' 2024.

\bibitem{zhong2020towards}
Y.~Zhong and W.~Deng, ``Towards transferable adversarial attack against deep face recognition,'' \emph{IEEE Transactions on Information Forensics and Security}, vol.~16, pp. 1452--1466, 2020.

\bibitem{komkov2021advhat}
S.~Komkov and A.~Petiushko, ``Advhat: Real-world adversarial attack on arcface face id system,'' in \emph{25th International Conference on Pattern Recognition (ICPR)}.\hskip 1em plus 0.5em minus 0.4em\relax IEEE, 2021, pp. 819--826.

\bibitem{xue2021backdoors}
M.~Xue, C.~He, J.~Wang, and W.~Liu, ``Backdoors hidden in facial features: A novel invisible backdoor attack against face recognition systems,'' \emph{Peer-to-Peer Networking and Applications}, vol.~14, pp. 1458--1474, 2021.

\bibitem{wenger2021backdoor}
E.~Wenger, J.~Passananti, A.~N. Bhagoji, Y.~Yao, H.~Zheng, and B.~Y. Zhao, ``Backdoor attacks against deep learning systems in the physical world,'' in \emph{Proceedings of the IEEE Conference on Computer Vision and Pattern Recognition}, 2021, pp. 6206--6215.

\bibitem{damer2018morgan}
N.~Damer, A.~M. Saladie, A.~Braun, and A.~Kuijper, ``Morgan: Recognition vulnerability and attack detectability of face morphing attacks created by generative adversarial network,'' in \emph{2018 IEEE 9th International Conference on Biometrics Theory, Applications and Systems}.\hskip 1em plus 0.5em minus 0.4em\relax IEEE, 2018, pp. 1--10.

\bibitem{scherhag2017vulnerability}
U.~Scherhag, R.~Raghavendra, K.~B. Raja, M.~Gomez-Barrero, C.~Rathgeb, and C.~Busch, ``On the vulnerability of face recognition systems towards morphed face attacks,'' in \emph{2017 5th International Workshop on Biometrics and Forensics (IWBF)}.\hskip 1em plus 0.5em minus 0.4em\relax IEEE, 2017, pp. 1--6.

\bibitem{taskiran2020face}
M.~Taskiran, N.~Kahraman, and C.~E. Erdem, ``Face recognition: Past, present and future (a review),'' \emph{Digital Signal Processing}, vol. 106, p. 102809, 2020.

\bibitem{Invixium}
Invixium, ``Invixium remote face enrollment service,'' \url{https://www.invixium.com/remote-face-enrollment/}.

\bibitem{HikVision}
HikVision, ``Hikvision face recognition terminals,'' \url{https://www.hikvision.com/en/products/Access-Control-Products/Face-Recognition-Terminals/}.

\bibitem{li2020light}
H.~Li, Y.~Wang, X.~Xie, Y.~Liu, S.~Wang, R.~Wan, L.-P. Chau, and A.~C. Kot, ``Light can hack your face! black-box backdoor attack on face recognition systems,'' \emph{arXiv preprint arXiv:2009.06996}, 2020.

\bibitem{venkatesh2021face}
S.~Venkatesh, R.~Ramachandra, K.~Raja, and C.~Busch, ``Face morphing attack generation and detection: A comprehensive survey,'' \emph{IEEE Transactions on Technology and Society}, vol.~2, no.~3, pp. 128--145, 2021.

\bibitem{wolberg1998image}
G.~Wolberg, ``Image morphing: a survey,'' \emph{The Visual Computer}, vol.~14, no. 8-9, pp. 360--372, 1998.

\bibitem{jia2022adv}
S.~Jia, B.~Yin, T.~Yao, S.~Ding, C.~Shen, X.~Yang, and C.~Ma, ``Adv-attribute: Inconspicuous and transferable adversarial attack on face recognition,'' \emph{Advances in Neural Information Processing Systems}, vol.~35, pp. 34\,136--34\,147, 2022.

\bibitem{amada2021universal}
T.~Amada, S.~P. Liew, K.~Kakizaki, and T.~Araki, ``Universal adversarial spoofing attacks against face recognition,'' in \emph{2021 IEEE International Joint Conference on Biometrics (IJCB)}.\hskip 1em plus 0.5em minus 0.4em\relax IEEE, 2021, pp. 1--7.

\bibitem{yang2020design}
X.~Yang, F.~Wei, H.~Zhang, and J.~Zhu, ``Design and interpretation of universal adversarial patches in face detection,'' in \emph{Computer Vision--ECCV 2020: 16th European Conference, Glasgow, UK, August 23--28, 2020, Proceedings, Part XVII 16}.\hskip 1em plus 0.5em minus 0.4em\relax Springer, 2020, pp. 174--191.

\bibitem{guo2021master}
W.~Guo, B.~Tondi, and M.~Barni, ``A master key backdoor for universal impersonation attack against dnn-based face verification,'' \emph{Pattern Recognition Letters}, vol. 144, pp. 61--67, 2021.

\bibitem{sarkar2020facehack}
E.~Sarkar, H.~Benkraouda, and M.~Maniatakos, ``Facehack: Triggering backdoored facial recognition systems using facial characteristics,'' \emph{arXiv preprint arXiv:2006.11623}, 2020.

\bibitem{deb2020advfaces}
D.~Deb, J.~Zhang, and A.~K. Jain, ``Advfaces: Adversarial face synthesis,'' in \emph{2020 IEEE International Joint Conference on Biometrics (IJCB)}.\hskip 1em plus 0.5em minus 0.4em\relax IEEE, 2020, pp. 1--10.

\bibitem{kurakin2018adversarial}
A.~Kurakin, I.~J. Goodfellow, and S.~Bengio, ``Adversarial examples in the physical world,'' in \emph{Artificial intelligence safety and security}.\hskip 1em plus 0.5em minus 0.4em\relax Chapman and Hall/CRC, 2018, pp. 99--112.

\bibitem{demontis2019adversarial}
A.~Demontis, M.~Melis, M.~Pintor, M.~Jagielski, B.~Biggio, A.~Oprea, C.~Nita-Rotaru, and F.~Roli, ``Why do adversarial attacks transfer? explaining transferability of evasion and poisoning attacks,'' in \emph{28th USENIX Security Symposium (USENIX Security)}, 2019, pp. 321--338.

\bibitem{huang2019enhancing}
Q.~Huang, I.~Katsman, H.~He, Z.~Gu, S.~Belongie, and S.-N. Lim, ``Enhancing adversarial example transferability with an intermediate level attack,'' in \emph{Proceedings of the IEEE International Conference on Computer Vision}, 2019, pp. 4733--4742.

\bibitem{zhou2018transferable}
W.~Zhou, X.~Hou, Y.~Chen, M.~Tang, X.~Huang, X.~Gan, and Y.~Yang, ``Transferable adversarial perturbations,'' in \emph{Proceedings of the European Conference on Computer Vision (ECCV)}, 2018, pp. 452--467.

\bibitem{zhao2019seeing}
Y.~Zhao, H.~Zhu, R.~Liang, Q.~Shen, S.~Zhang, and K.~Chen, ``Seeing isn't believing: Towards more robust adversarial attack against real world object detectors,'' in \emph{Proceedings of the 2019 ACM SIGSAC Conference on Computer and Communications Security}, 2019, pp. 1989--2004.

\bibitem{xu2020adversarial}
K.~Xu, G.~Zhang, S.~Liu, Q.~Fan, M.~Sun, H.~Chen, P.-Y. Chen, Y.~Wang, and X.~Lin, ``Adversarial t-shirt! evading person detectors in a physical world,'' in \emph{Computer Vision--ECCV 2020: 16th European Conference, Glasgow, UK, August 23--28, 2020, Proceedings, Part V 16}.\hskip 1em plus 0.5em minus 0.4em\relax Springer, 2020, pp. 665--681.

\bibitem{wiyatno2019physical}
R.~R. Wiyatno and A.~Xu, ``Physical adversarial textures that fool visual object tracking,'' in \emph{Proceedings of the IEEE International Conference on Computer Vision}, 2019, pp. 4822--4831.

\bibitem{meng2021magface}
Q.~Meng, S.~Zhao, Z.~Huang, and F.~Zhou, ``Magface: A universal representation for face recognition and quality assessment,'' in \emph{Proceedings of the IEEE Conference on Computer Vision and Pattern Recognition}, 2021, pp. 14\,225--14\,234.

\bibitem{wu2023face}
H.~Wu, V.~Albiero, K.~Krishnapriya, M.~C. King, and K.~W. Bowyer, ``Face recognition accuracy across demographics: Shining a light into the problem,'' in \emph{Proceedings of the IEEE Conference on Computer Vision and Pattern Recognition}, 2023, pp. 1041--1050.

\bibitem{sharif2016accessorize}
M.~Sharif, S.~Bhagavatula, L.~Bauer, and M.~K. Reiter, ``Accessorize to a crime: Real and stealthy attacks on state-of-the-art face recognition,'' in \emph{Proceedings of the 2016 ACM SIGSAC Conference on Computer and Communications Security}, 2016, pp. 1528--1540.

\bibitem{wei2023physically}
X.~Wei, J.~Yu, and Y.~Huang, ``Physically adversarial infrared patches with learnable shapes and locations,'' in \emph{Proceedings of the IEEE Conference on Computer Vision and Pattern Recognition}, 2023, pp. 12\,334--12\,342.

\bibitem{karras2017progressive}
T.~Karras, T.~Aila, S.~Laine, and J.~Lehtinen, ``Progressive growing of gans for improved quality, stability, and variation,'' \emph{arXiv preprint arXiv:1710.10196}, 2017.

\bibitem{chen2018mobilefacenets}
S.~Chen, Y.~Liu, X.~Gao, and Z.~Han, ``Mobilefacenets: Efficient cnns for accurate real-time face verification on mobile devices,'' in \emph{Biometric Recognition: 13th Chinese Conference, CCBR 2018, Urumqi, China, August 11-12, 2018, Proceedings 13}.\hskip 1em plus 0.5em minus 0.4em\relax Springer, 2018, pp. 428--438.

\bibitem{zhang2018unreasonable}
R.~Zhang, P.~Isola, A.~A. Efros, E.~Shechtman, and O.~Wang, ``The unreasonable effectiveness of deep features as a perceptual metric,'' in \emph{Proceedings of the IEEE Conference on Computer Vision and Pattern Recognition}, 2018, pp. 586--595.

\bibitem{kingma2014adam}
D.~P. Kingma and J.~Ba, ``Adam: A method for stochastic optimization,'' \emph{arXiv preprint arXiv:1412.6980}, 2014.

\bibitem{chen2023rethinking}
H.~Chen, Y.~Zhang, Y.~Dong, X.~Yang, H.~Su, and J.~Zhu, ``Rethinking model ensemble in transfer-based adversarial attacks,'' in \emph{The Twelfth International Conference on Learning Representations}, 2023.

\bibitem{he2016deep}
K.~He, X.~Zhang, S.~Ren, and J.~Sun, ``Deep residual learning for image recognition,'' in \emph{Proceedings of the IEEE Conference on Computer Vision and Pattern Recognition}, 2016, pp. 770--778.

\bibitem{deng2019arcface}
J.~Deng, J.~Guo, N.~Xue, and S.~Zafeiriou, ``Arcface: Additive angular margin loss for deep face recognition,'' in \emph{Proceedings of the IEEE Conference on Computer Vision and pattern Recognition}, 2019, pp. 4690--4699.

\bibitem{wang2018cosface}
H.~Wang, Y.~Wang, Z.~Zhou, X.~Ji, D.~Gong, J.~Zhou, Z.~Li, and W.~Liu, ``Cosface: Large margin cosine loss for deep face recognition,'' in \emph{Proceedings of the IEEE Conference on Computer Vision and Pattern Recognition}, 2018, pp. 5265--5274.

\bibitem{liu2017sphereface}
W.~Liu, Y.~Wen, Z.~Yu, M.~Li, B.~Raj, and L.~Song, ``Sphereface: Deep hypersphere embedding for face recognition,'' in \emph{Proceedings of the IEEE Conference on Computer Vision and Pattern Recognition}, 2017, pp. 212--220.

\bibitem{riba2020kornia}
E.~Riba, D.~Mishkin, D.~Ponsa, E.~Rublee, and G.~Bradski, ``Kornia: an open source differentiable computer vision library for pytorch,'' in \emph{Proceedings of the IEEE Winter Conference on Applications of Computer Vision}, 2020, pp. 3674--3683.

\bibitem{huang2008labeled}
G.~B. Huang, M.~Mattar, T.~Berg, and E.~Learned-Miller, ``Labeled faces in the wild: A database forstudying face recognition in unconstrained environments,'' in \emph{Workshop on Faces in Real-Life Images: Detection, Alignment, and Recognition}, 2008.

\bibitem{zheng2017cross}
T.~Zheng, W.~Deng, and J.~Hu, ``Cross-age lfw: A database for studying cross-age face recognition in unconstrained environments,'' \emph{arXiv preprint arXiv:1708.08197}, 2017.

\bibitem{moschoglou2017agedb}
S.~Moschoglou, A.~Papaioannou, C.~Sagonas, J.~Deng, I.~Kotsia, and S.~Zafeiriou, ``Agedb: the first manually collected, in-the-wild age database,'' in \emph{Proceedings of the IEEE Conference on Computer Vision and Pattern Recognition Workshops}, 2017, pp. 51--59.

\bibitem{zhang2016joint}
K.~Zhang, Z.~Zhang, Z.~Li, and Y.~Qiao, ``Joint face detection and alignment using multitask cascaded convolutional networks,'' \emph{IEEE Signal Processing Letters}, vol.~23, no.~10, pp. 1499--1503, 2016.

\bibitem{rombach2022high}
R.~Rombach, A.~Blattmann, D.~Lorenz, P.~Esser, and B.~Ommer, ``High-resolution image synthesis with latent diffusion models,'' in \emph{Proceedings of the IEEE Conference on Computer Vision and Pattern Recognition}, 2022, pp. 10\,684--10\,695.

\bibitem{wang2019neural}
B.~Wang, Y.~Yao, S.~Shan, H.~Li, B.~Viswanath, H.~Zheng, and B.~Y. Zhao, ``Neural cleanse: Identifying and mitigating backdoor attacks in neural networks,'' in \emph{2019 IEEE Symposium on Security and Privacy (SP)}.\hskip 1em plus 0.5em minus 0.4em\relax IEEE, 2019, pp. 707--723.

\bibitem{gao2019strip}
Y.~Gao, C.~Xu, D.~Wang, S.~Chen, D.~C. Ranasinghe, and S.~Nepal, ``Strip: A defence against trojan attacks on deep neural networks,'' in \emph{Proceedings of the 35th Annual Computer Security Applications Conference}, 2019, pp. 113--125.

\bibitem{liu2019abs}
Y.~Liu, W.-C. Lee, G.~Tao, S.~Ma, Y.~Aafer, and X.~Zhang, ``Abs: Scanning neural networks for back-doors by artificial brain stimulation,'' in \emph{Proceedings of the 2019 ACM SIGSAC Conference on Computer and Communications Security}, 2019, pp. 1265--1282.

\bibitem{zhu2022toward}
Y.~Zhu, Y.~Chen, X.~Li, K.~Chen, Y.~He, X.~Tian, B.~Zheng, Y.~Chen, and Q.~Huang, ``Toward understanding and boosting adversarial transferability from a distribution perspective,'' \emph{IEEE Transactions on Image Processing}, vol.~31, pp. 6487--6501, 2022.

\end{thebibliography}

\appendix
\setcounter{table}{0}   
\setcounter{figure}{0}
\setcounter{section}{0}
\renewcommand{\thetable}{\arabic{table}}
\renewcommand{\thefigure}{\arabic{figure}}
\newcommand{\alphsubsection}[1]{\renewcommand{\thesubsection}{\arabic{subsection}}\subsection{#1}}

\begin{figure}[h]
    \centering
    \includegraphics[width=0.6\linewidth]{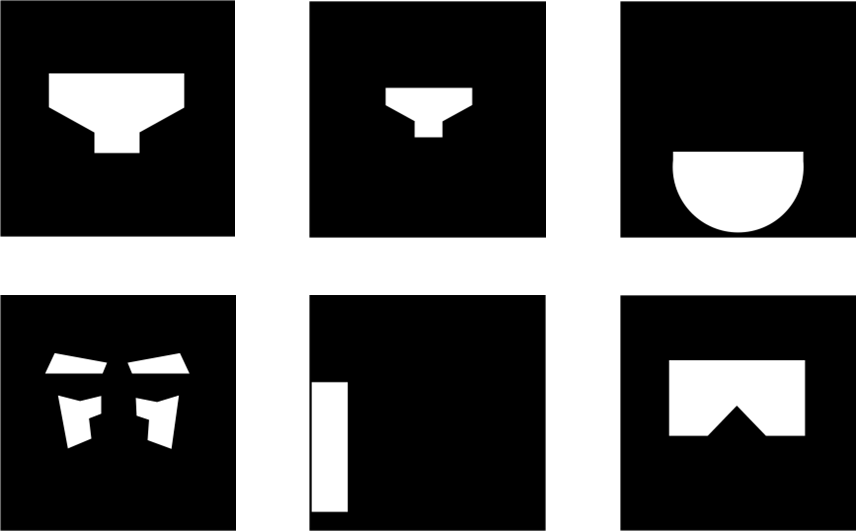}
    \caption{All mask types used in our experiments.}
    \label{fig:mask_all}
\end{figure}

\alphsubsection{Mask of Different Facial Regions}
\label{appendix:mask}
Various mask types are used in previous studies, including mouth, nose, cheek, eyebrow and forehead. To measure the different attack performance of different masks and how these facial regions matter in face recognition, the masks listed in Figure \ref{fig:mask_all} are used.

\alphsubsection{Key Feature Mask Searching}
\label{appendix:algorithm}
To find key feature mask for a given facial image using a facial feature extractor model, we devise a key feature mask searching algorithm shown Algorithm \ref{alg:mask}. It initializes a random mask matrix and the base feature embedding of the input image. Then, it iterates over a specified number of optimization steps. In each step, it generates an adversarial image by combining the original image with the current mask and random noise. It calculates three loss components: TV loss to encourage spatial smoothness, edge loss to discourage sharp edges in the mask, and a cosine loss to maximize the difference between the adversarial image's features and the base features. The mask is updated by taking a signed gradient descent step with respect to the combined loss. Finally, the mask is clipped to the range [0, 1] and smoothed using a median blur filter.

\begin{algorithm}[h]
\renewcommand{\algorithmicrequire}{\textbf{Input:}}
\renewcommand{\algorithmicensure}{\textbf{Output:}}
\caption{Key Feature Mask Searching}
\label{alg:mask}
\begin{algorithmic}[1] 
\REQUIRE  $x$: facial image; $f_{\theta}$: facial feature extractor with $\theta$; $K$: optimization step, $cr$: cover rate of the facical image.
\ENSURE  $m$: feature mask
Set $m$ to random matrix $\mathcal{N}(0,1)$\\
Set $e_b$ to $f_{\theta}(x)$ \\
\FOR {$k = 0:K$}
    \STATE $x_{adv}=x\cdot(1-m)+m\cdot \mathcal{N}(0,1)$
    \STATE $\mathcal{L}_{tv} = \sum_{i, j}\left(\left(m_{i, j}-m_{i+1, j}\right)^2+\left(m_{i, j}-m_{i, j+1}\right)^2\right)^{\frac{1}{2}}$
    \STATE $\mathcal{L}_{edge} = \|\Psi(m)\|_2$
    \STATE $\mathcal{L}_{all} = \mathcal{L}_{tv} + \mathcal{L}_{edge} - \mathrm{cos}(f_{\theta}(x_{adv}),e_b)$
    \STATE $m=m-\alpha\cdot\mathrm{sign}(\frac{\nabla_{m}\mathcal{L}_{all}}{\|\nabla_{m}\mathcal{L}_{all}\|_2})$
    \STATE $m = \mathrm{Clip}(\text{{medianBlur}}(m, 3),0,1)$
    \STATE $m_{i,j} = \begin{cases}
    0, & \text{if } mask_{i,j} \leq \mathrm{Quantile}(m, 1 - \text{cr}) \\
    1, & \text{otherwise}
    \end{cases}$
\ENDFOR
\STATE \textbf{Return} $m$
\end{algorithmic}
\end{algorithm}

\alphsubsection{Backdoor Trigger Generation of FIBA}
As shown in Algortithm \ref{alg:fiba}, FIBA aims to generate a backdoor trigger for facial images. The input consists of a set of feature extractors, a set of facial images for trigger training, a facial image of the insider, a selected mask, a set of transformations, and the steps for optimization.

The algorithm initializes the backdoor trigger by multiplying the target insider's facial image with the selected mask. Then, it enters a loop for the specified number of optimization steps. In each iteration, the algorithm applies transformations to the target image combined with the current backdoor trigger. It calculates the loss function, which comprises three components: total variation loss, LPIPS loss, and edge loss. Additionally, it subtracts the similarity loss between the transformed and training images for each feature extractor. The backdoor trigger is updated using the gradients of the loss function using the Adam optimizer.

\begin{algorithm}[h]
\renewcommand{\algorithmicrequire}{\textbf{Input:}}
\renewcommand{\algorithmicensure}{\textbf{Output:}}
\caption{FIBA: Backdoor Trigger Generation}
\label{alg:fiba}
\begin{algorithmic}[1] 
\REQUIRE  $\mathbb{M}$: set of facial feature extractors; $x_{train}$: facial images collected to train the backdoor trigger; $x_v$: facial image of the insider; $m$: seleted mask to optimize backdoor trigger; $\mathbb{T}$: transformation set. $N$: steps to optimize backdoor trigger.
\ENSURE  $p$: generated backdoor trigger.
\STATE Set $p$ to $x_v\cdot m$ \\
\FOR {$n = 0:N$}
    \STATE $\hat{x}=\mathbb{T}(x_v\cdot (1-m) + m\cdot p)$
    \STATE $\mathcal{L} = \alpha\mathcal{L}_{tv} + \gamma\mathcal{L}_{lpips} + \beta\mathcal{L}_{egdge} $
    \FOR {$f_{\theta}^{i}$ in $\mathbb{F}$}
        \STATE $\mathcal{L} -= \mathcal{L}_{sim}(f_{\theta}^{i},\hat{x},x_{train})$
    \ENDFOR
    \STATE Update $p$ with $\nabla_{p}\mathcal{L}$ using $\mathrm{Adam}$ optimizer
\ENDFOR
\STATE \textbf{Return} $p$
\end{algorithmic}
\end{algorithm}

\alphsubsection{Defense Strategy Against FIBA}
\label{appendix:defense}
Defense strategy against FIBA to enhance the robustness of a facial feature extractor model, is given in Algorithm \ref{alg:defense}. It takes a facial feature extractor $f_{\theta}$, a dataset $\mathcal{D}{ft}$ for fine-tuning, and a linear layer $h{\theta^{\prime}}$ to map the embedding to the number of classes as input. The algorithm iterates over the dataset and generates adversarial examples by adding perturbations to the input images. The perturbations are calculated using the gradient ascent to maximize the loss function $\mathcal{L}{adv}$, which encourages the embeddings of different identities to be similar. The final loss function $\mathcal{L}$ combines $\mathcal{L}{adv}$ with the cross-entropy loss for the original classification task, allowing the model to learn both robust facial features and accurate classification simultaneously.

\begin{algorithm}[h]
\renewcommand{\algorithmicrequire}{\textbf{Input:}}
\renewcommand{\algorithmicensure}{\textbf{Output:}}
\caption{Defense Strategy Against FIBA}
\label{alg:defense}
\begin{algorithmic}[1] 
\REQUIRE  $f_{\theta}$: facial feature extractor with $\theta$; $\mathcal{D}_{ft}$: dataset to finetue the model; $h_{\theta^{\prime}}$: linear layer to map embedding to number of class.
\ENSURE  $f_{\theta}$: finetuned model.
\FOR {$n = 0:N$}
    \FOR {$(x,y)$ in $\mathcal{D}_{ft}$}
        \STATE Initialize $\delta$ with $\mathcal{N}(0,1)$ \\
        \STATE $\hat{e} = f_{\theta}(x+\delta)$ \\
        \STATE $\mathcal{L}_{adv} = \sum_{i,j} (\mathrm{1}-\mathrm{E})\odot (\hat{e}\cdot \hat{e}^{T})$ \\
        \STATE $\delta=\delta+\alpha\cdot\mathrm{sign}(\frac{\nabla_{\delta}\mathcal{L}_{adv}}{\|\nabla_{\delta}\mathcal{L}_{adv}\|_2})$ \\
        \STATE $\hat{e} = f_{\theta}(x+\delta)$ \\
        \STATE $\mathcal{L} = \sum_{i,j} (\mathrm{1}-\mathrm{E})\odot (\hat{e}\cdot \hat{e}^{T}) + \gamma\cdot\mathrm{CE}(h(f_{\theta}(x)),y)$ \\
        \STATE Update $\theta$ with $\nabla_{\theta}\mathcal{L}$
    \ENDFOR
\ENDFOR
\STATE \textbf{Return} $f_{\theta}.$
\end{algorithmic}
\end{algorithm}

\begin{figure}[h]
    \centering
    \includegraphics[width=1\linewidth]{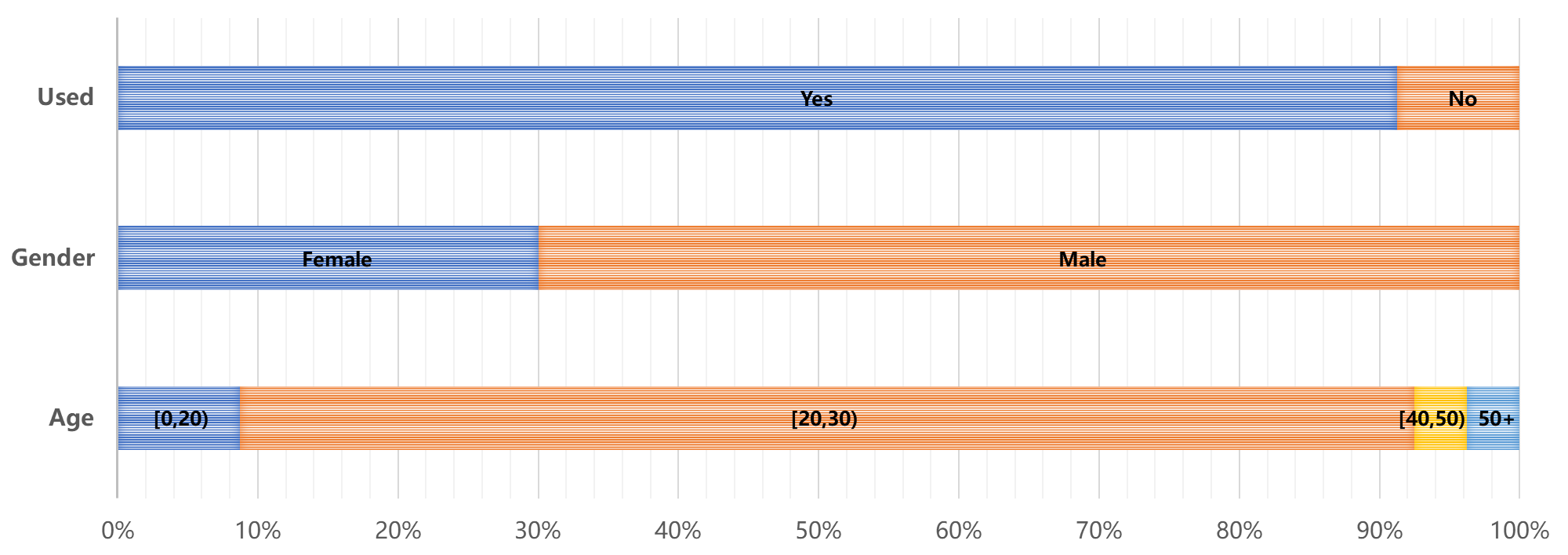}
    \caption{The distribution of the surveyed users.}
    \label{fig:user-info}
\end{figure}

\alphsubsection{Details of User Study}
\label{appendix:user_study}
Using a questionnaire, we have conducted a user study involving 80 participants (70\% male and 30\% female), among which 73 volunteers (91.25\%) have used FRS before. The volunteers recruited are most university students around us (83.75\% of them aged from 20 to 30). Details are shown in Figure \ref{fig:user-info}.

\alphsubsection{Details of Datasets}
\label{appendix:datasets}
In the experiments, we use several datasets including: LFW \cite{huang2008labeled}, CALFW \cite{zheng2017cross}, AgeDB \cite{moschoglou2017agedb}, and CelebA-HQ \cite{karras2017progressive}. The LFW (Labeled Faces in the Wild) dataset is widely recognized as a benchmark for face recognition tasks, consisting of 13,233 labeled face images of 5,749 individuals collected from the internet. The CALFW (Cross-Age LFW) dataset serves as an extension to the LFW dataset, constructed by deliberately selecting 3,000 positive face pairs with age gaps to incorporate the aging process's intra-class variance. The AgeDB dataset contains 16,488 images of various famous individuals. The CelebA-HQ dataset is a high-quality version of the CelebA dataset, which comprise 30,000 images of 6,217 unique identities, each with a resolution of 1024×1024. 

\alphsubsection{Details of FIBA}
\label{appendix:fiba}
For proposed attack FIBA, we use Adam optimizer with learning rate 100 and the weight of Total Variation (TV) loss, lpips loss and edge loss are set to 0.1, 0.05 and 0.001. For each generation, the iterations of the optimization is set to 200. Besides, we initialize the patch using one of the facial images of the insider's images of the same region for better fit and faster convergence. The detailed parameters of the patch transformation module are presented in Table \ref{tab:parameters}.

\begin{table}[h]
\centering
\caption{Parameters of Transformation.}
\label{tab:parameters}
\begin{tabular}{ll}
\toprule
Transformation & Parameters \\
\midrule
ColorJiggle & brightness\_factor=0.15, \\
            & brightness\_range=[0.65, 0.9], \\
            & max\_delta=0.1, probability=1.0 \\
\midrule
RandomAffine & translate=(0.05, 0.05), \\
             & probability=1.0, degrees=5, \\
\midrule
RandomGaussianBlur & kernel\_size=(3, 3), \\
                   & sigma\_range=(0.1, 2.0), \\
                   & probability=0.4 \\
\bottomrule
\end{tabular}
\end{table}

\alphsubsection{Details of IoT Devices}
\label{appendix:devices}
We have evaluated FIBA on three IoT devices from three different service vendors. All of them support FLV to avoid face forge. They use different way to enroll the facial images including digital and physical domain. The detailed ID and other information of the devices are presented in Table \ref{tab:device_config}.

\begin{table}[t]
\centering
\caption{Configuration of IoT devices.}
\label{tab:device_config}
\small
\renewcommand{\arraystretch}{0.95}
\begin{tabular}{ccccc}
\hline
ID & Brand & Model ID & FLV & Enrollment \\ \hline
D-1 & HIKVISION & A10 & Yes & Taking Photos \\
D-2 & MoreDian & MY5003A & Yes & Taking Photos \\
D-3 & SenseTime & S7 & Yes & Uploading Photos \\ \hline
\end{tabular}
\end{table}

\alphsubsection{Details of Facial Feature Extractors}
\label{appendix:models}
We have evaluated FIBA on six popular face recognition models in digital experiments. These models are trained with different datasets and have different threshold to identify two facial images as the same identity. The detailed information of the models is presented in Table \ref{tab:model_info}.

\begin{table}[]
    \centering
    \caption{Backbone networks of target models.}
    \label{tab:model_info}
    \small
    \renewcommand{\arraystretch}{0.95}
    \begin{tabular}{cccc}
    \hline Target Models & Backbone & Training Dataset & Threshold \\
    \hline MobileFace & MobileFacenet & MS-Celeb-1M & 0.22 \\
    MobilenetV2 & MobileNetV2 & MS-Celeb-1M & 0.22 \\
    ShuffleNetV1 & ShuffleNetV1 & MS-Celeb-1M & 0.22 \\
    IR50-Softmax & IRSE-50 & MS-Celeb-1M & 0.26 \\
    IR50-SphereFace & IRSE-50 & MS-Celeb-1M & 0.33 \\
    CASIA-Softmax & IRSE-50 & CASIA & 0.33 \\
    \hline
    \end{tabular}
\end{table}

\begin{figure}[h]
  \centering
  \includegraphics[width=0.7\linewidth]{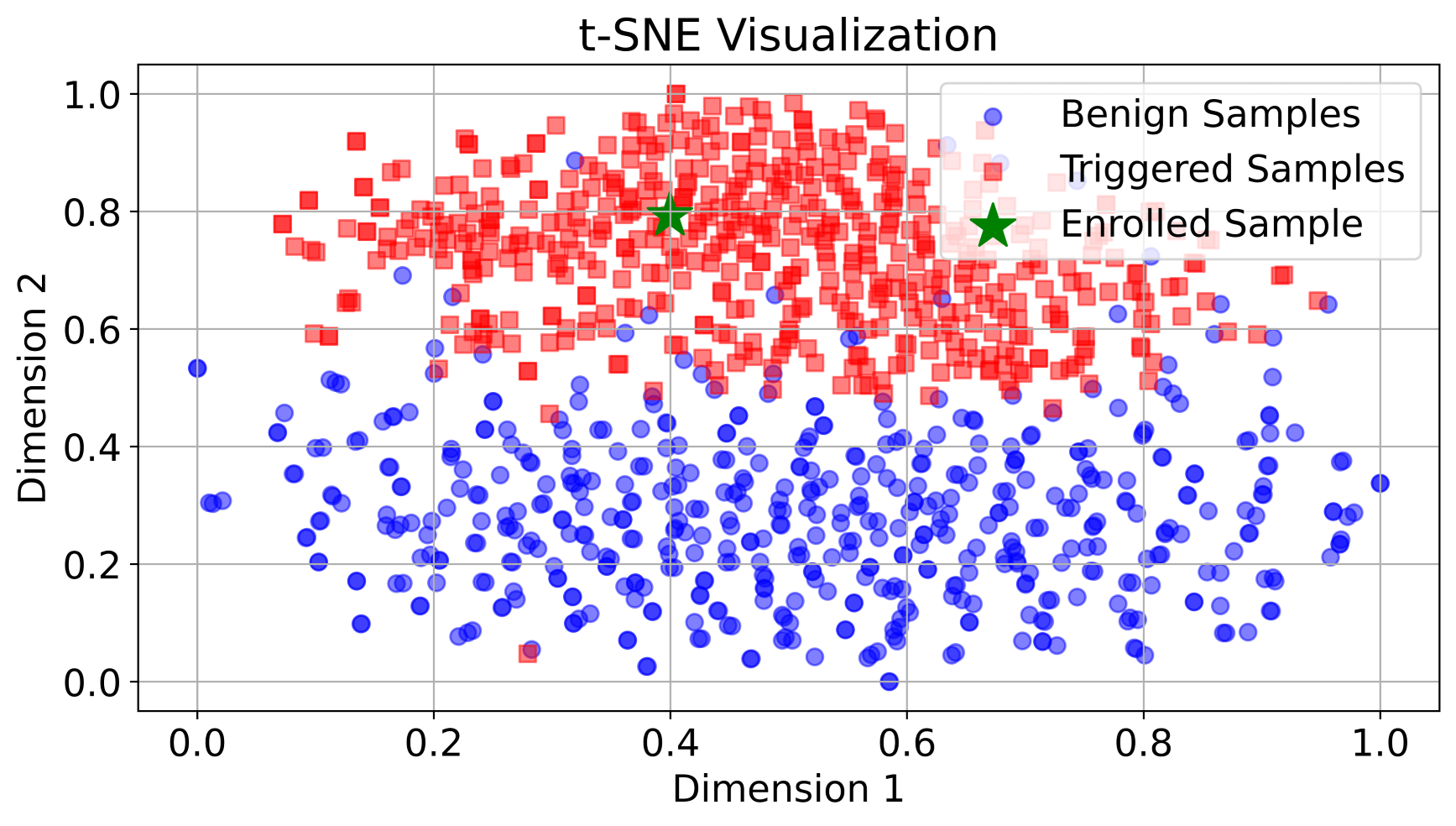}
  \caption{t-SNE visualization.}
  \label{fig:tsne}
\end{figure}

\alphsubsection{Visualization of the Feature Embedding}
\label{appendix:visualization}
As mentioned before, we visualize the features of the benign, triggered, and enrolled sample in Figure \ref{fig:tsne}. Upon applying t-SNE, we can see that as the benign face image is masked with a backdoor trigger, its features intuitively move towards that of the enrolled sample. This means that backdoor trigger can effectively replace their original feature with that of the trigger. This visualization also illustrates that even without the accessibility to training data of the feature extractor, FIBA can achieve similar performance with high generalization to different FRS as well.

\begin{figure}[h]
    \centering
    \includegraphics[width=0.7\linewidth]{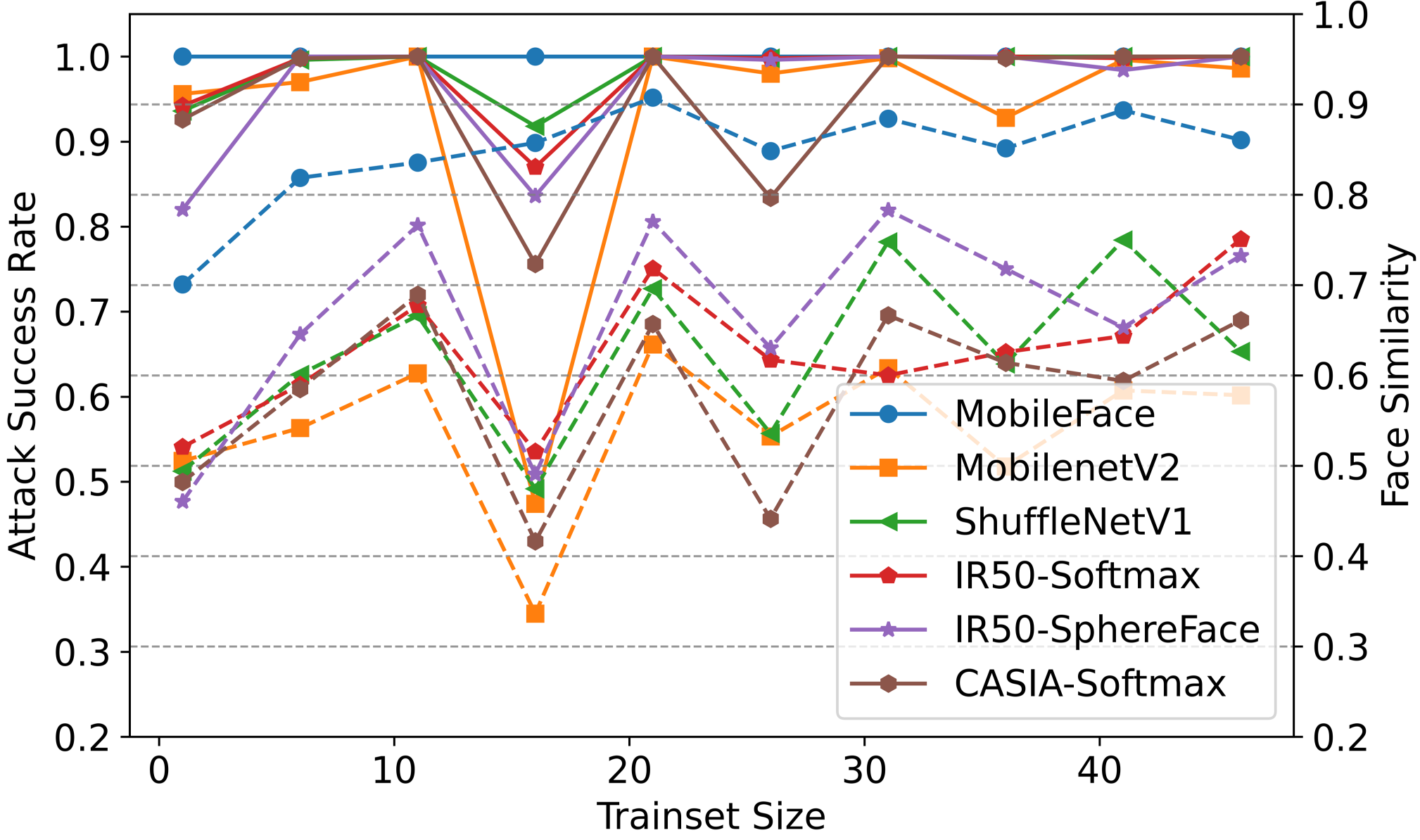}
    \caption{Impact of training data size on FIBA.}
    \label{fig:train_num}
\end{figure}

\alphsubsection{Number of the Training Data}
\label{appendix:train_size}
Note that in previous experiments, we the number of the training data is set to 50. However, how will this variable influences the performance of FIBA is still unknown and will this affect the transferability of FIBA. To figure out the above questions, we use MobileFace as a surrogate model and different number of images from CelebA as a training set to measure this factor. The generated patch are tested under both white-box and black-box settings using 500 images. The results are given in Figure \ref{fig:train_num} (The solid line refers to ASR, and the dashed line refers to face similarity). As we can see from the figure, the size of training data does not have a large impact on the performance of FIBA since FIBA can achieve more than 90\% ASR on most models. However, with the increasing size of training data, ASR and face similarity across different models become more and more stable with slight increase of model performance.

\begin{figure}[h]
    \centering
    \includegraphics[width=0.6\linewidth]{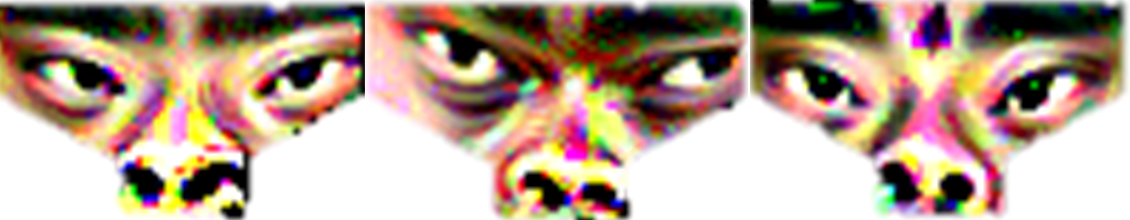}
    \caption{Examples of digital patch.}
    \label{fig:digital_patch}
\end{figure}

\begin{figure}[h]
    \centering
    \includegraphics[width=0.6\linewidth]{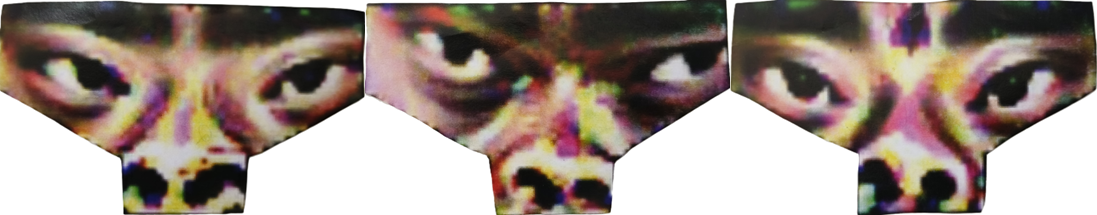}
    \caption{Examples of physical patch.}
    \label{fig:physcial_patch}
\end{figure}

\begin{figure}[h]
    \centering
    \includegraphics[width=0.6\linewidth]{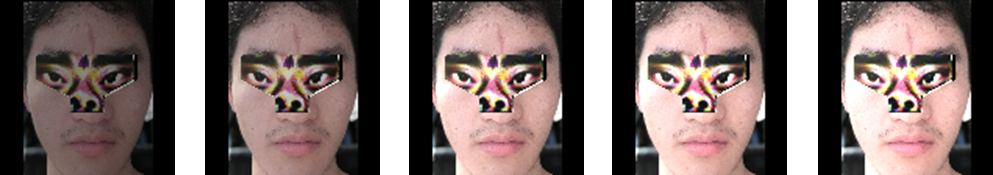}
    \caption{Evaluation under different light conditions.}
    \label{fig:example_light}
\end{figure}

\begin{figure}[h]
    \centering
    \includegraphics[width=0.6\linewidth]{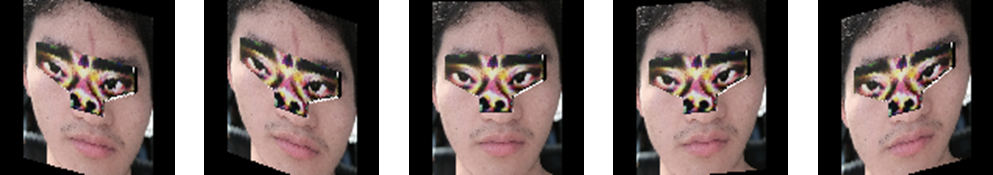}
    \caption{Evaluation with different shooting angles.}
    \label{fig:example_angles}
\end{figure}

\begin{figure}[h]
    \centering
    \includegraphics[width=0.6\linewidth]{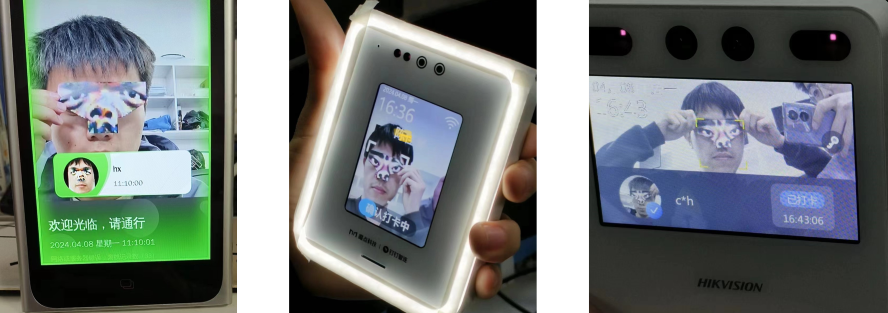}
    \caption{Examples for evaluating Three FRS IoT devices.}
    \label{fig:iot_example}
\end{figure}

\alphsubsection{Examples of Digital and Physical Mask}
\label{appendix:physical}
In this paper, to launch attack in physcial world, we first need to generate the backdoor trigger (or known as patch) in digital domain as shown in Figure \ref{fig:digital_patch}. Then, the digital mask will be printed using a 2D printer, as shown in Figure \ref{fig:physcial_patch}. For simplicity, we use the printed patch to test the impact of different light conditions and shooting angles on performance of FIBA. Specifically, we directly paste the printed patch on the facial images of the volunteers taken in physical domain. Note that this will not affect the evaluation of FIBA since physical distortions have been considered. As shown in Figure \ref{fig:example_light}, for different light conditions, we use python to adjust the light intensity of each image to control it accurately and avoid other factors that may happen in real-world scenarios. As shown in Figure \ref{fig:example_angles}, for different shooting angles, we use affine function to adjust the horizontal angles of the facial images. We have also provide the examples of IoT devices evaluation in Figure \ref{fig:iot_example}.

\cleardoublepage

\end{document}